\documentclass[epj]{svjour}
\usepackage{amssymb}
\usepackage{amsmath}
\usepackage{epsfig}
\usepackage{graphicx}% Include figure files
\DeclareGraphicsExtensions{.eps,.png,.jpg,.pdf}
\usepackage{svg}
\usepackage{bm}% bold math
\usepackage{kantlipsum}
\usepackage{multirow}% http://ctan.org/pkg/multirow
\usepackage{hhline}% http://ctan.org/pkg/hhline
\usepackage{textcomp}% http://ctan.org/pkg/hhline
\usepackage{url}
\begin{document}

%\preprint{APS/123-QED}

\title{Radiative neutron capture cross section of $^{242}$Pu measured at n\_TOF-EAR1 in the unresolved resonance region up to 600~keV}% Force line breaks with \\

\author{%
J.~Lerendegui-Marco\inst{1,2} \and %
C.~Guerrero\inst{1} \and %
E.~Mendoza\inst{3} \and %
J.~M.~Quesada\inst{1} \and %
K.~Eberhardt\inst{4}\and
A.R. Junghans\inst{5} \and %
V. Alcayne\inst{3} \and %
V. Babiano\inst{2} \and %
O.~Aberle\inst{6} \and %
J.~Andrzejewski\inst{7}
% ...
\and %
the n\_TOF Collaboration
}
\institute{
Dpto. F\'isica At\'omica, Molecular y Nuclear, Universidad de Sevilla, Spain \and
Instituto de F\'isica Corpuscular, CSIC - Universidad de Valencia, Spain \and
Centro de Investigaciones Energ\'{e}ticas Medioambientales y Tecnol\'{o}gicas (CIEMAT), Spain \and
Johannes Gutenberg-Universit¨at Mainz, 55128 Mainz, Germany \and
Helmholtz-Zentrum Dresden-Rossendorf, D-01328 Dresden, Germany \and
European Organization for Nuclear Research (CERN), Switzerland \and
University of Lodz, Poland
% ...
} 

\date{Received: \today / Revised version: \today}

\abstract{Accurate neutron capture cross sections are essential for the design and operation of fast reactors using MOX fuels. For $^{242}$Pu, the Nuclear Energy Agency (NEA) recommends 8--12\% accuracy in the fast energy region (2--500~keV), compared to the current uncertainty of 35\%. Moreover, integral experiments and previous measurements suggest the evaluated $^{242}$Pu(n,$\gamma$) cross section is overestimated, particularly in the JEFF-3.3 library, which shows a 14\% overestimation between 1~keV and 1 MeV. Recent measurements from LANSCE reported a 20--30\% reduction in the 1--40~keV range relative to evaluations. To solve these discrepancies, the $^{242}$Pu(n,$\gamma$) cross section was measured from 1 to 600~keV at CERN n\_TOF-EAR1 facility using a 95(4)~mg $^{242}$Pu target, enriched to 99.959\%. Gamma rays from neutron capture were detected with an array of C$_6$D$_6$ scintillators and a novel application of the Pulse Height Weighting Technique was employed. The resulting cross section presents a systematic uncertainty between 8 and 12\%, reducing the current uncertainties of 35\% and achieving the accuracy requested by the NEA. Analysis using FITACS produced average resonance parameters, consistent with the analysis of the resolved resonance region. Our data align well with Wisshak and Kaeppeler, and are 10--14\% lower than JEFF-3.3 in the 1--250~keV range, helping to achieve consistency with integral benchmarks. At higher energies, our results are in reasonable agreement with ENDF\/B-VIII.1 and JEFF-3.3. In contrast, DANCE results appear to underestimate the cross section by a factor of 2--3 above a few~keV.
\PACS{
      {PACS-key}{discribing text of that key}   \and
      {PACS-key}{discribing text of that key}
     } % end of PACS codes
}%end of abstract

\maketitle

\begingroup
\let\clearpage\relax
\twocolumn
\endgroup

%\tableofcontents

\section{\label{sec:Intro} Introduction}
The design and operation of current and innovative nuclear systems such as Accelerator Driven
Systems (ADS) and Gen-IV reactors, aimed at improving the sustainability of nuclear energy, requires accurate neutron cross sections
of relevant isotopes~\cite{1:proposal,WPEC26}. These
advanced nuclear systems are expected to work with different fuel compositions, such as MOX~\cite{IAEA_MOX}, and in other neutron energy regimes than current thermal power reactors. Thus the interaction of
neutrons of different energies with the isotopes present in the new fuels must be investigated in detail. Indeed, most of these systems (see Table~\ref{Table_summary})
are fast reactors, and the 
reduction of cross section uncertainties in the~keV range becomes crucial~\cite{WPEC26}.

This is also the case of neutron induced reactions on $^{242}$Pu and especially of its capture cross section. Table~\ref{Table_summary} summarizes the current and required accuracies
in the $^{242}$Pu($n$,$\gamma$) cross sections in the energy range of interest for different nuclear systems. The first attempts to measure the $^{242}$Pu($n$,$\gamma$) reaction in its unresolved resonance region (URR)
were made in 1975 when Hockenbury et al.~\cite{Hockenbury:1975} ca-

\hfill \break
\hfill \break
\hfill \break
\hfill \break
\hfill \break
\hfill \break
\hfill \break
\hfill \break
\hfill \break
\hfill \break
\hfill \break
\hfill \break
\hfill \break
\hfill \break
\hfill \break
\hfill \break
\hfill \break
\hfill \break
\hfill \break
\hfill \break
\hfill \break
\hfill \break
\hfill \break
\hfill \break
\hfill \break
\hfill \break
\hfill \break
\hfill \break
\hfill \break
\hfill \break
\hfill \break
\hfill \break
\hfill \break
\hfill \break
\hfill \break
\hfill \break
\hfill \break
rried out a time-of-flight measurement (6--87~keV) at the Rensselaer Polytechnic Institute (RPI). 

A few years later, Wisshak and K\"appeler~\cite{Wisshak_Kaeppeler:1978,Wisshak_Kaeppeler:1979} measured the $^{242}$Pu capture cross section relative to that of $^{197}$Au 
in the unresolved resonance region in two energy intervals, 10 to 90~keV and 50 to 250~keV, at the Forschungzentrum Karlsruhe (FZK). 
Based on these data sets, the NEA WPEC-26 group estimates that the current uncertainty of this cross section ranges from 24 to 39\%~\cite{WPEC26} in the
neutron energy range $E_n$=2--500~keV.

\begin{figure}[!b]
\begin{center}
\includegraphics[width=9.0cm]{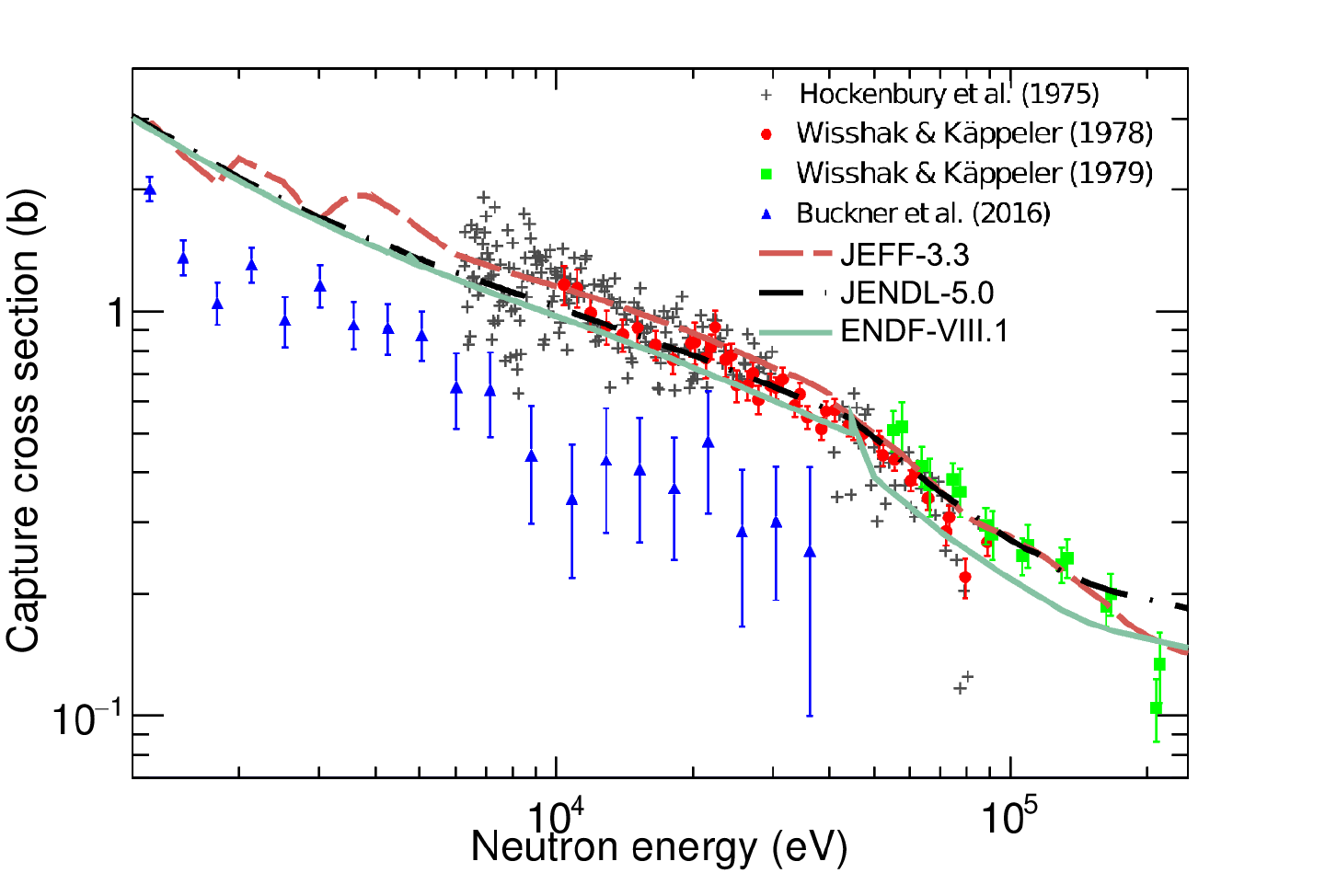}
%\includesvg[width=9.0cm]{Fig0_Pu242_ng_Previous_measurements_evaluations_intro_legend_fullnames_Cut1keV_updatedLibraries.svg}
%\includegraphics[width=9.0cm]{Fig0_Pu242_ng_Previous_measurements_evaluations_intro_legend_fullnames}
\caption{Capture cross section of $^{242}$Pu in the energy range from 1 to 250~keV. The previous experimental data sets available in EXFOR are compared to the cross section
reported in the main evaluated files.}
\label{Fig_StatusURR}
\end{center}
\end{figure}

Recently, another time-of-flight measurement 
was carried out with the DANCE detector at LANSCE by Buckner et al.~\cite{Buckner:2016}, covering the region from thermal to 40~keV. Their result 
suggests a systematic reduction of 20--30\% in the URR (above 1~keV) compared to the evaluated cross section of ENDF/B-VIII.1~\cite{ENDF}. A summary of the main features of all these measurements is presented in Table~\ref{Table_PrevMeas}.
Among the evaluations, JEFF-3.3~\cite{JEFF} is in agreement with ENDF/B-VIII.1 around 1~keV but 10--20\% above the latter at $\approx$5--30~keV, while JENDL-5.0~\cite{JENDL} is intermediate between JEFF and ENDF up to 50~keV and is in agreement or above JEFF beyond this energy. Fig.~\ref{Fig_StatusURR} illustrates the current status of the $^{242}$Pu($n$,$\gamma$) cross section in terms of evaluations and experimental data in the fast energy region. 

\begin{table}
\begin{center}
\caption{Current and required accuracy in the $^{242}$Pu($n$,$\gamma$) cross section for nuclear innovative systems (fast reactors) according to Ref.~\cite{WPEC26}. The range of accuracy represent the maximum and minimum values among the different neutron energy (E$_{n}$) groups.
SFR: Sodium Cooled Fast Reactor, EFR: European Fast Reactor, GFR: Gas Cooled Fast Reactor, LFR: Lead Cooled Fast Reactor, ADMAB: Accelerator-Driven Minor Actinide Burner, ADS: Accelerator-Driven System.}
\begin{tabular}{ccccc} 
\hline \hline
	                    & &&\multicolumn{2}{c}{\textbf{Accuracy(\%)}} \\ 
          & \textbf{E$_{n}$ Range} && \textbf{Current} &  \textbf{Required} \\
\hline
SFR                 &    2--500~keV    &&      24--39       &  8--12     \\
EFR                 &    2--67~keV    &&      $\sim$35     &  25-28    \\
GFR                 &    2--183~keV    &&      $\sim$35    &  8--13     \\
LFR                 &    9--183~keV    &&      $\sim$35    &  12     \\
ADMAB (ADS)          &    9--25~keV     &&      $\sim$35    & 10    \\
\hline \hline    
\end{tabular} 
\label{Table_summary}
\end{center}
\end{table}

\begin{table*}[!t]
\begin{center}
\caption{Main features of the previous time-of-flight measurements of the $^{242}$Pu($n$,$\gamma$) cross section in the URR compiled in EXFOR~\cite{EXFOR}.}
\begin{tabular}{ccccc} 
\hline \hline
 \textbf{Ref.}       & \textbf{Facility/Detector}       & \textbf{Final data}        & \textbf{E$_{n}$ Range} & \textbf{Unc. (\%)}  \\
\hline
Hockenbury et al. (1975)~\cite{Hockenbury:1975}     &  RPI/Liquid scintillation    &   Absolute cross section                      &      6--87~keV                  &    n.a.         \\
Wisshak and K\"appeler (1978)~\cite{Wisshak_Kaeppeler:1978}  &  FZK/Moxon-Rae               &   Ratio to $^{197}$Au($n$,$\gamma$)               &      10--90~keV                 &    6--12                    \\
Wisshak and K\"appeler (1979)~\cite{Wisshak_Kaeppeler:1979}  &  FZK/Moxon-Rae               &   Ratio to $^{197}$Au($n$,$\gamma$)               &      50--250~keV                &    10--20                        \\
Buckner et al. (2016)~\cite{Buckner:2016}             &  LANSCE/Total absorption     &   Absolute cross section                      &      1--40~keV           &    6--61                       \\
\hline \hline    
\end{tabular} 
\label{Table_PrevMeas}
\end{center}
\end{table*}

In this context, the NEA WPEC-26 Subgroup recommends in Ref.~\cite{WPEC26} that the capture cross section of $^{242}$Pu should be measured with an improved accuracy in the $E_n$=2--500~keV range. The precise requirements in 
terms of energy range and target uncertainty depend on the specific system (see Table~\ref{Table_summary}). 

Furthermore, simulations with JEFF-3.3 of the PROFIL and PROFIL-2 post irradiation experiments carried out in the fast reactor PHENIX concluded that $^{242}$Pu shows the largest deviation 
between the calculated and experimental capture rates among all the studied isotopes. The results, on average, indicate that the JEFF evaluation overestimates the $^{242}$Pu integral capture cross section in the region between 1~keV and 1 MeV
by 14\%~\cite{Noguere:2005,Tommasi:2006,Tommasi:2008}. For this reason, the NEA High Priority Request List (HPRL)~\cite{HPRL} endorsed also  request for 
new high-resolution measurements of the $^{242}$Pu($n$,$\gamma$) cross section
in the resonance region (0.5 eV to 2~keV) in order to obtain accurate average 
resonance parameters aiming at the reevaluation of the fast energy region. 

Following the requirements of the NEA, a new time-of-flight measurement of the capture cross section of $^{242}$Pu was carried out in 2015 in the experimental area 1 (EAR1) of the n\_TOF facility and preliminary results were presented in Refs.~\cite{JLWonder15,JLND2016}. 
The first publication on this measurement~\cite{JLerendegui:RRR} dealt with the analysis of the resonance
region and the statistical properties of resonance parameters. The resonance region was analyzed up to 4~keV with a systematic uncertainty of 5\%, addressing the requirements
of the NEA-HPRL. Individual and average resonance parameters were obtained from the \textit{R}-Matrix analysis of 250 resonances. An additional measurement using the same $^{242}$Pu targets addressed the thermal cross Sec.~\cite{JLerendegui:Thermal}.

This paper focuses on the analysis of the URR ($E_n>1$~keV), where a careful study of the different background contributions becomes crucial. In the following section we describe briefly the n\_TOF-EAR1 facility and the experimental set-up. 
The data reduction to obtain capture yield, focusing on the background determination, and the limitations to expand the energy range above 600~keV, is described in Sec.~\ref{sec:Analysis}.
Last, the validation of the analysis using the $^{197}$Au($n$,$\gamma$) ancillary measurement and the comparison of results to the evaluations and existing data
are discussed in Section \ref{sec:Results} and the conclusions are presented in Sec.~\ref{sec:Summary}.

\section{\label{sec:Measurement} Measurement at \lowercase{n}\_TOF}

\subsection{\label{sec:nTOF} The n\_TOF facility at CERN}

The neutron beam at n\_TOF is generated through spallation of lead nuclei induced by 20~GeV/c protons extracted in pulses from the CERN Proton Synchrotron and impinging (at the time of this measurement) on a cylindrical lead target 40~cm
in length and 60~cm in diameter. These pulses feature a nominal intensity of $7 \times 10^{12}$ protons, delivered with a time spread of 7~ns (rms) at a maximum frequency of 
0.83~Hz. The resulting high energy (MeV--GeV) spallation neutrons are partially moderated in a surrounding water layer to produce a white-spectrum neutron beam that expands in energy from thermal to a few GeV. At the time of the measurements,
neutrons travelled along two beam lines towards two experimental areas: EAR1 at 185~m (horizontal) \cite{Guerrero:2013} and EAR2 at 19~m (vertical) \cite{Weiss:2015} aimed at time-of-flight (TOF) experiments. The newest addition to n\_TOF is the recently built NEAR activation station~\cite{Patronis:24}, at
a distance of only 2.5~m from the lead spallation target. Each of the TOF experimental areas is better suited for certain types of measurements, depending on the specific requirements in terms of flux and resolution. While EAR1 features a better time-of-flight 
(i.e. neutron energy) resolution, the shorter vertical beam line of EAR2 provides a 400 times higher instantaneous neutron flux (see Ref.~\cite{JLerendegui:2016}), which makes it specially well suited for measuring 
highly radioactive and/or small mass samples (see for instance Refs. \cite{Barbagallo:Be7,Balibrea:NPA,Lerendegui:NPA}). The measurement of the $^{242}$Pu($n$,$\gamma$) reaction at n\_TOF, aiming at describing accurately 
both the resolved and unresolved resonance regions up to the highest possible energy, is not affected by a high radioactivity background, and was therefore performed at n\_TOF-EAR1.
This is the first capture measurement on $^{242}$Pu performed at the n\_TOF facility, where other neutron induced 
reactions on Pu isotopes have been previously studied~\cite{Guerrero:Pu240,Tsinganis:Pu240,Stamatopoulos:Pu242}.

\subsection{\label{sec:Detection} Targets, detectors and operation during the measurement}

\begin{figure}[!h]
\begin{center}
\includegraphics[width=9.0cm]{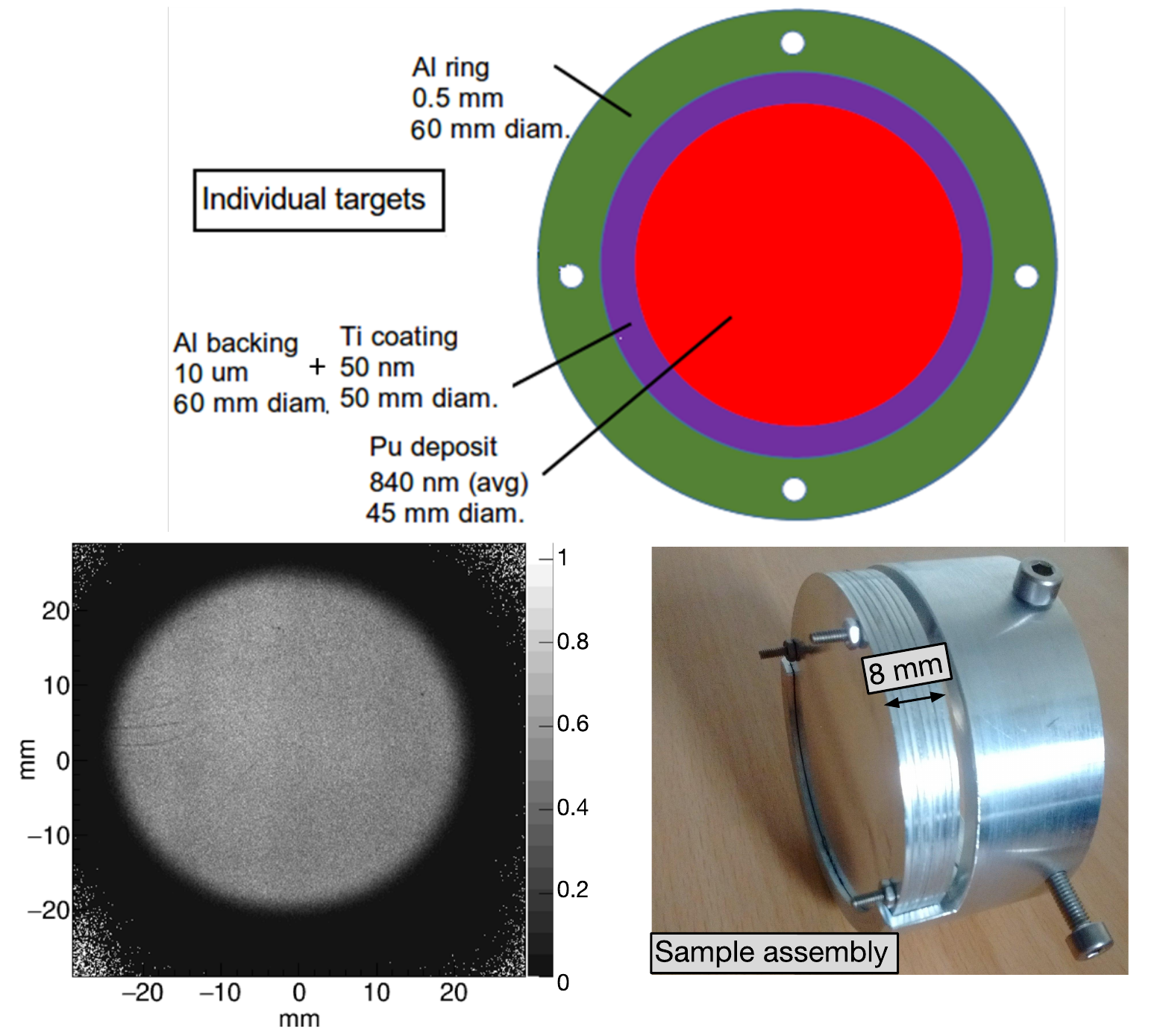}
\caption{{Top: Sketch of one of the targets ($^{242}$Pu on a thin Al backing)}. Bottom left: $\alpha$-radiography of one of the targets showing the homogeneity of the $^{242}$Pu deposits.
Bottom right: Mounted stack of $^{242}$Pu targets placed in a purpose built holder.}
\label{Fig_sample}
\end{center}
\end{figure}

The target preparation was carried out within the CHANDA project \cite{CHANDA} by the JGU University Mainz and the HZDR research center using 99.959$\%$ pure $^{242}$Pu provided and characterized by ORNL. A total of 95(4) mg of $^{242}$Pu were electrodeposited on seven thin ($\sim$10~$\mu$m) Al backings, each of them coated with a 50~nm thick Ti layer. 
The homogeneity in thickness for all seven targets combined was found to be $<$0.1\% in alpha-radiography measurements (see Fig.~\ref{Fig_sample}).
The seven thin backings were assembled in a stack of targets with a total thickness of 8~mm (out of which 7~mm is air). The main impurities, 2$\times$10$^{-4}$ of $^{240}$Pu and 5$\times$10$^{-5}$ of $^{239}$Pu, respectively, lead to a negligible ($\leq$0.1\%) background induced by capture and fission. More details on the $^{242}$Pu target preparation and composition can be found in Refs.~\cite{Eberhardt:2017,Guerrero:2018,JLerendegui:RRR}.

One particular aspect of this measurement is the use of several thin fission-like targets instead of a single thick target.
The $^{242}$Pu targets designed for this measurement, featuring each an average density of 0.85~mg/cm$^{2}$ with 45~mm diameter on thin Al backings, present several advantages 
with respect to the typical thick targets used in ($n$,$\gamma$) measurements~\cite{Guerrero:2018}:
\begin{itemize}
\item Higher mass than most of the previous capture measurements on minor actinides at n\_TOF~(see for instance Refs. \cite{237Np,241Am,243Am}).
\item High target to backing mass ratio.
\item Lower background arising from capture and scattering in the backing than in previous capture measurements.
\end{itemize}

%The associated Figure of Merit $FOM$ can be expressed as
%\begin{equation}
% FOM~=~\frac{n_{Pu242}~\cdot~\sigma^{MACS}_{Pu242}}{\sum_{i} n_{i} \cdot \sigma^{MACS}_{i}},
%\end{equation}
%$n_{Pu242}$ and $n_{i}$ being the surface density of $^{242}$Pu and the i-component of the backing, respectively, and $\sigma^{MACS}$
%are the corresponding maxwellian averaged total cross sections at kT = 30~keV. 
%Increasing this value is key to perform a succesful capture cross section measurement in the URR, where scattering is 2-3 orders above capture. 
Indeed, the backing-related neutron-induced background was the main limitation to extend
the analyzable neutron energy range in previous measurements at n\_TOF. Moreover, the large dimension of the $^{242}$Pu 
deposits in our target ensures that the full neutron beam is intercepted and thus no correction for a possible misalignment of the target is required. 
Lastly, the use of a large and thin $^{242}$Pu target also present advantages in the 
implementation of the Pulse Height Weighting Technique~\cite{Domingo:2006} (see Sec.~\ref{sec:Efficiency})
and in the calculation of the average cross section in the URR, 
due to the negligible self-absorption and multiple scattering corrections (see Sec.~\ref{sec:Results}). 
%\begin{figure}[!b]
%\begin{center}
%\epsfig{file=242Pu_sample_assembly_modified_7targets_removed_smallLetters.png, width=5.9cm}
%\includegraphics[width=5.5cm]{Fig2_Setup_sampleassembly_placed_vertical.png}
%\caption{Top: The four C$_{6}$D$_{6}$ scintillation detectors used to measure the capture $\gamma$-rays in n\_TOF-EAR1. Bottom: View of the $^{242}$Pu 7-target assembly placed in the neutron beam.}
%\label{Fig_setup}
%\end{center}
%\end{figure}

Ancillary measurements were carried out in this experimental campaign for background estimation, normalization and validation of the result in the URR. 
An exact replica of the backings assembly without
$^{242}$Pu deposits, called the \textit{dummy} target hereafter, was used to assess the beam-related background not associated with interactions with the $^{242}$Pu targets. 
The neutron energy dependence of the background due to neutrons and in-beam $\gamma$-rays scattered on the $^{242}$Pu
targets was estimated from the measurement of a Pb (1~mm thick) target (see Sec.~\ref{sec:Backgrounds}). Last, the measurement of a 100~$\mu$m thick $^{197}$Au target
with the same diameter as the $^{242}$Pu one was used for normalization using the Saturated Resonance Method (SRM)~\cite{SRM_Macklin}. 
In addition, the $^{197}$Au($n$,$\gamma$) cross section is a standard 
above 200~keV~\cite{IAEA:Standards} and is known with high accuracy in 
the energy region between 5 and 80~keV as well~\cite{Lederer:2011,Massimi:2014}. Therefore, it was used  for validation of the analysis of the $^{242}$Pu data in the URR (see Sec.~\ref{sec:Aucrosssection}). 

%\begin{itemize}
%\item Brief about sample or directly refer to RRR paper? 
%These first to items copy and rephrase/sum up from EAR1's paper 
%\item highlight the importance of having fast detectors with low mass and very low neutron sensitivity to go high in energy in the URR (gamma flash and elastic x-section 2-3 orders above capture)
%1\item highlight the importance of thin backgings and compare to previous cases (Np-237, 240-Pu, Am-241, Am-243)  (Include figure like Chanda's talk?) where the thick backings limited the energy region that
%could be analyzed.
%\end{itemize}

%Capture measurements at n\_TOF have been carried out to date using two different systems with different methodologies. 
%and the $\gamma$-ray multiplicity (i.e. number of fired crystals).

A set of four C$_{6}$D$_{6}$ Detectors~\cite{Plag} was used for detection of $\gamma$ rays. Their fast recovery 
from the so-called $\gamma$-flash (i.e. prompt $\gamma$-rays and ultra-relativistic particles produced in the spallation reactions), allows one
to measure up to a neutron energy of at least 1~MeV. Moreover, the extremely low neutron sensitivity is a clear advantage in the URR, where the scattering cross section clearly dominates over the capture one. The set-up of four detectors (see Fig.~2 of Ref.~\cite{JLerendegui:RRR}) is placed upstream from the $^{242}$Pu target to minimize the in-beam $\gamma$-ray background and at an angle of 
125$^{\circ}$ with respect to the neutron beam. This placement minimizes the impact of anisotropic emission
of the primary $\gamma$-rays for capture events with $l>0$.

The proton and neutron beam intensities were monitored using a Wall current monitor that measures the proton beam current, 
and an array of four silicon detectors with a $^{6}$Li neutron converter~\cite{Marrone:2004}. 
These systems provided compatible results within 0.5\% for all the measurements in the campaign.
%setting the accuracy in the scaling when comparing or combining different measurements, for instance for background subtraction and normalization. 
More details on the monitoring system are given in the 
previous $^{242}$Pu($n$,$\gamma$) publication~\cite{JLerendegui:RRR}. The CERN Proton Synchrotron provided during this measurement high intensity pulses ($\sim 7 \times 10^{12}$ protons) together with low intensity ones ($\sim 2\times10^{12}$ protons). 
The latter were not used to extract the cross section but served to validate the behavior of the detector response in the more demanding conditions of the high intensity pulses, and the 
corresponding higher $\gamma$-flash and counting rate.

%additional detectors are used at n\_TOF for monitoring the proton and neutron beams. 
%A Wall Current Monitor (so-called PKUP) measures the intensity of the proton beam in each pulse.
%The neutron beam is monitored with the SiMon system \cite{Marrone:2004}, an array of four 
%silicon detectors placed outside the beam and looking at a thin enriched lithium fluoride foil for detecting the products of the $^{6}$Li(n,$\alpha$) reaction,
%whose cross section is considered a standard from thermal energy to 1~MeV. 
%The pulse-by-pulse intensities provided by these two detectors have been found to be proportional within 0.5\%,
%which sets the accuracy in the scaling when comparing or combining different measurements, for instance for background subtraction and normalization.

The n\_TOF Data Acquisition system~\cite{DAQpaper}, consisting of 14-bits flash ADCs, is triggered with the arrival of each proton pulse to the neutron-producing 
target and records the detector
output signals during a time-of-flight range of 100~ms (down to 18~meV in neutron energy)
for both capture and monitoring detectors. The full electronic signals are automatically transferred from the Data Acquisition (DAQ) computers to the
CERN Advanced STORage manager (CASTOR) for their long-term storage and off-line analysis.

%Each of the detectors in the capture set-up and the beam monitoring system is connected to one channel of the n\_TOF Data acquistion system
%which features 12-bit digitizers sampling at 900~MSamples/s during 100~ms following the arrival of the proton pulse to the spallation target 
%(i.e. recording signals of reactions induced by neutrons for energies down to 18 meV). The full data movies are automatically transferred from the DAQ00
%computers to the CERN Advanced STORage manager (CASTOR) for their long-term storage and offline analysis.

\section{\label{sec:Analysis} Analysis}

\subsection{\label{sec:Calibrations} Amplitude and time-of-flight calibrations}

The waveforms for each detector are processed with a Pulse Shape Analysis routine~\cite{Zugec:PSA}
that extracts the amplitude, area and time-of-flight of each signal
together with the information on the proton pulse (date, time, type and intensity). In the second step, histograms of counts as a function 
of the time-of-flight and amplitude are built for each individual detector and measured sample.

Calibrations of signal amplitude to deposited $\gamma$-ray energy were performed on a 
weekly basis to correct for possible gain shifts during the 1-month-long measurement
using $^{137}$Cs, $^{88}$Y, $^{241}$Am/$^{9}$Be and $^{244}$Cm/$^{13}$C calibration sources.  The uncertainty in the capture cross section associated with the impact of these gain  shifts in the weighted spectra (see Sec.~\ref{sec:Efficiency}) has been estimated to be only 0.5\%. More details can be found in Refs.~\cite{JLerendegui:RRR,JLerendegui:Thesis}.

\begin{figure}[!h]
\begin{center}
\includegraphics[width=8.5cm]{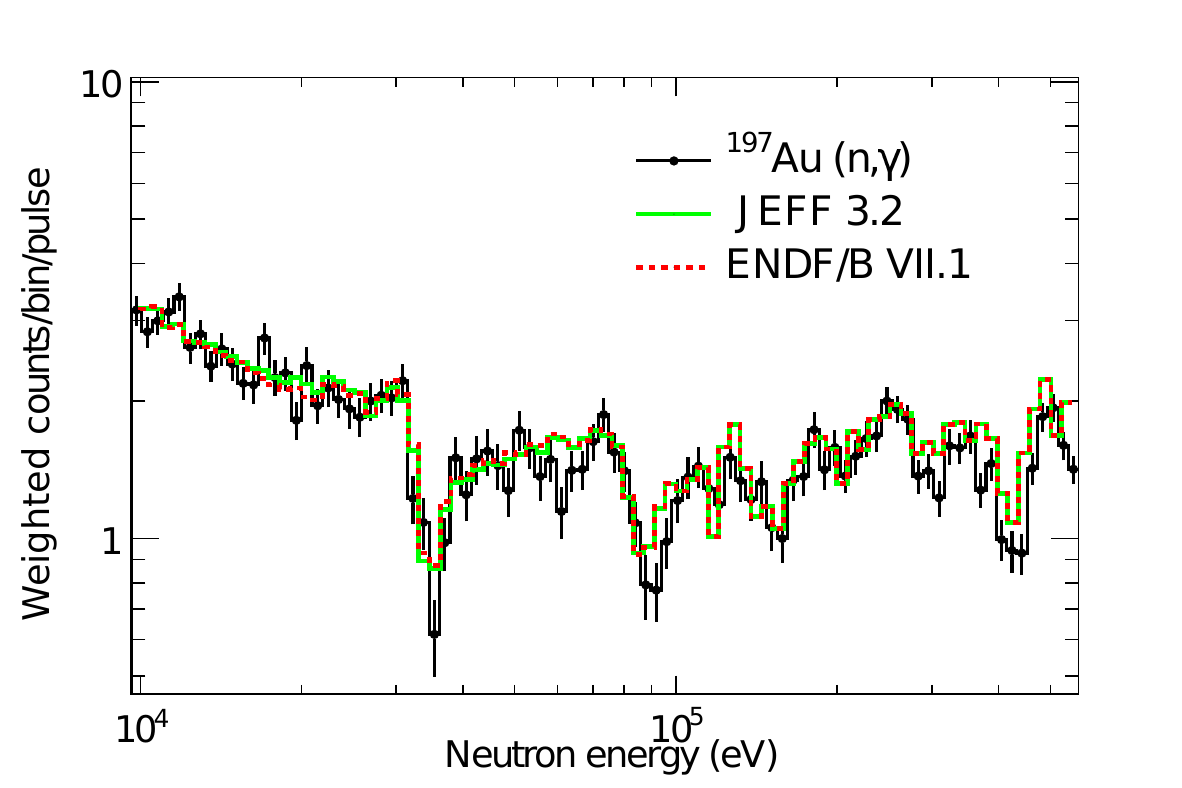}
\caption{Experimental weighted counting rate of Au as a function of the reconstructed neutron energy compared to the ENDF and JEFF cross sections convoluted with the n\_TOF evaluated flux~\cite{Barbagallo:2013}. The energies of the main dips in the flux have been used to fit $t_{off}=100$ ns (see text for details).}
\label{Fig_t2e_Audips}
\end{center}
\end{figure}

The time-to-energy calibration in the URR was carried out by matching
the energy position of the most prominent absorption dips (corresponding to resonances of Al and Mn present in the spallation target exit window) in the n\_TOF-EAR1 evaluated flux $\phi(E_{n})$~\cite{Barbagallo:2013}, 
in the energy region between 30 and 500~keV, to their corresponding position in the time-of-flight distribution of
\begin{equation}
E_{n}=\frac{1}{2}m_{n}(\frac{L_{0}}{t_{m}-t_0-t_{\rm off}})^{2},
\label{t2e}
\end{equation}
where $m_{n}$ is the neutron mass, $t_m$ is the arrival time of a neutron determined from the detection of the reaction products, $t_0 = t_{\gamma} - L_{0}/c$ is the start time, calculated from the arrival time of the $\gamma$-flash to
each detector (t$_{\gamma}$) and the time-of-flight of $\gamma$-rays from the spallation target to EAR1 ($L_0/c$). 
$L_{0}$=183.88(5)~m is the effective flightpath that provides a good energy calibration at low neutron energies (4.9 eV $^{197}$Au resonance from Ref.~\cite{JEFF}). 
The time offset $t_{\rm off}$ in Eq.~(\ref{t2e}) is required to take into account the non-univocal relation between the energy of a neutron and its arrival time to EAR1,
the so-called Resolution Function of the facility~\cite{Lorusso:2004,Guerrero:2013,LoMeo:2015}. Following the approach described in Ref.~\cite{Lorusso:2004}, a value of $t_{\rm off}$=-100(30)~ns was fitted to match the neutron energy of the dips in the flux with the corresponding time-of-flight in the $^{197}$Au($n$,$\gamma$) data. The good reproduction of the dip energies in the flux using this time-to-energy calibration is shown in Fig.~\ref{Fig_t2e_Audips}, where we compare the neutron-energy-calibrated $^{197}$Au($n$,$\gamma$) counting rate with the expected counting rate $\sigma_{\gamma}(E_{n}) \times \phi(E_{n})$ according to ENDF/B-VIII.1 and JEFF-3.3. 

%A time offset $T_{off}$ = -100(30)~ns was extracted from the fit of the energy and the time of the flux dips in Fig.~\ref{Fig_t2e_Audips}. 

%\begin{equation} 
%E_{n}=\frac{1}{2}m_{n}(\frac{L_{0}+\lambda (E_{n}) }{t_{m}-t_0})^{2},
%\label{eq:t2e}
%\end{equation}
%where $m_{n}$ is the neutron mass, $t_0 = t_{\gamma} - L/c$ is the start time, calculated from the arrival time of the $\gamma$-flash to
%each detector (t$_{\gamma}$) and $L/c$, the time-of-flight of $\gamma$-rays from the target to EAR1. $\lambda (E_{n})$ is an energy-dependent equivalent moderation length, related to the 
%Resolution Function of the facility, that was extracted from GEANT4 simulations of the n\_TOF-EAR1 neutron beam~\cite{LoMeo:2015}. Last, $L_{0}$ is the flight path from the  
%reference position at which $\lambda (E_{n})$ was calculated to the experimental area and was found to be to $L_{0}$=183.73(5)~m using the 4.9 eV resonance of gold as reference.The reader
%is referred to Refs.~\cite{Lorusso:2004,Guerrero:2013, LoMeo:2015} for a more detailed explanation of the Resolution Function. 

%The time-to-energy calibration in the URR should perfectly match the dips in the counting rate with the ones in the n\_TOF-EAR1 evaluated flux~\cite{Barbagallo:2013}
%(absorption in Al and Mn in the spallation target) to avoid non-physical fluctuations in the resulting capture yield.
%Aiming at matching the dips in the neutron flux and in the capture counting rate, the time-of-flight (t-t$_{\gamma}$) position of the clear dips in the $^{197}$Au($n$,$\gamma$) counting rate was calibrated to match the corresponding

%in the was obtained from the fit. 
%The fair reproduction of the position of the dips after this effective calibration is applied is shown in the bottom pad of Fig.~\ref{Fig_t2e_Audips},
%where we compare the expected counting rate (evaluated cross section times the 
%neutron flux) with the background substracted $^{197}$Au($n$,$\gamma$) counting rate as a function of the calibrated neutron energy. 

%Indeed, the shape of the flux in this region is known within 4-5\% and 
%becomes a relevant systematic uncertainty as it is discussed later in this paper.

%\begin{itemize}
%\item Brief on signal analysis, event building and histogramming
%\item Brief description of the TED + refer to details from the paper on the RRR
%\item Time to energy relation: Pragmatic approach to match flux and dips
%\end{itemize}

\subsection{\label{sec:Efficiency} Detection efficiency: Modified PHWT and neutron-energy dependency}
In general, the detection efficiency for a capture cascade depends on its multiplicity and the energy of the individual $\gamma$-rays that, in general, depend on the neutron resonance.
To eliminate this dependence, neutron capture measurements with C$_{6}$D$_{6}$ detectors are traditionally following the Total Energy Detection (TED) concept~\cite{TED} in combination with Pulse Height Weighting Technique (PHWT)~\cite{Tain:PHWT} that gives different weight to each detected signal energy via so-called Weighting function (WF). Following the prescription of Ref.~\cite{Tain:PHWT}, the WF (analytical functions, typically $3^{rd}$--$4^{th}$ degree polynomials) for the measured $^{242}$Pu and $^{197}$Au samples have been calculated from the detector response to 16 mono-energetic $\gamma$-rays from 50~keV to 10~MeV obtained via Monte Carlo simulations of the detection system
carried out with the Geant4 toolkit~\cite{Geant4_1,Geant4_2} (see Figure~2 in Ref.~\cite{JLerendegui:RRR}). It is important to remark that the simplicity of this analytical approach for the PHWT is limited to the use of thin targets~\cite{Guerrero:2018}. For the case of thick targets a more complex numerical WF is required~\cite{Domingo:2006}.

%This is based on two conditions:
%\begin{enumerate}
%\item The efficiency of the detectors must be low enough so that at most one $\gamma$-ray per cascade is detected. 
%In this case the efficiency for detecting a cascade $\varepsilon_{c}$ becomes 
%\begin{equation} 
%\varepsilon_{c}=1-\prod_{i}(1-\varepsilon_{i})\approx\sum_{i}\varepsilon_{i},
%\label{eq:TED1}
%\end{equation} 
%$\varepsilon_{i}$ being the efficiency to detect the i$^{th}$ $\gamma$-ray of the cascade. 
%The sum goes over all the $\gamma$-rays emitted after a capture event.

%\item The efficiency to detect each $\gamma$-ray is proportional to its energy $E_{i}$, hence Eq.~(\ref{eq:TED1}) develops into
%\begin{equation} 
%\varepsilon_{c} = \sum_{i}k E_{i}= k \sum_{i} E_{i}= k E_{c}. 
%\label{eq:TED3}
%\end{equation} 
%\end{enumerate}

%Under these two conditions the efficiency for detecting a cascade is proportional to the known initial excitation energy of the compound nucleus (E$_{c}$) and independent of the actual decay pattern.
%However, the second requirement is not fulfilled by C$_{6}$D$_{6}$ detectors and therefore a mathematical manipulation of the detector response is needed to  achieve the necessary proportionality between detection efficiency and $\gamma$-ray energy: the counts recorded at each deposited energy are weighted by a factor dependent on its energy (pulse height), given by the so-called  weighting function (WF). This is known as the \textit{Pulse Height Weighting Technique (PHWT)}. 

The use of the PHWT leads in most cases to a reduction of the background if composed by low energy $\gamma$-rays, as the corresponding weight is lower than for capture events. However, in this work this is not the case because the dominant background, measured with the \textit{dummy} target (see Sec.~\ref{sec:Backgrounds}), presents a more energetic 
$\gamma$-ray spectrum than that of capture on $^{242}$Pu (see Ref.~\cite{JLerendegui:RRR}). Another effect of the PHWT that is usually neglected is that enhanced fluctuations appear in the energy bins with limited statistics~\cite{JLerendegui:RRR,Mendoza:23}. To avoid these two unwanted effects of the PHWT in our data, we have applied the PHWT following an alternative approach based on an average weighting factor (AWF):
\begin{enumerate}
 \item The background subtraction (see Sec.~\ref{sec:Backgrounds}) is carried out using the unweighted count distributions.
 \item The weighted and unweighted $^{242}$Pu($n$,$\gamma$) histograms are used to calculate the average weighting factor $\langle W^{thr}\rangle$ (i.e. the average ratio of weighted to unweighted resonance areas). This has been calculated, as in Ref.~\cite{JLerendegui:RRR}, using the s-wave resonances at E$_n\leq$1~keV.
   \item The capture yield is calculated from the background-subtracted unweighted counting distribution, which is then scaled using the average weighting factor (see Eqs.~(\ref{eq:Yieldexp}) and (\ref{eq:YieldNorm}) in Sec.~\ref{sec:Yield}). 
   \item After scaling with $\langle W^{thr}\rangle$, the efficiency to detect a cascade $\varepsilon_c$ in the calculation of the yield then would correspond to $\varepsilon_{c}^{thr} (E_n=0\  {\rm~keV})=S_{n}$, where $S_{n}$ corresponds is the neutron separation energy of the compound nucleus.
   \item The dependency of the efficiency with the neutron energy $\varepsilon_c^{thr}(E_n)$ in the URR has been calculated on the basis of cascade simulations using unweighted counts, as explained in the following.
\end{enumerate}

The AWF method was first used for the analysis of the resolved resonance region (RRR), where we proved that decay patterns do not change significantly among resonances due to very high level density in $^{243}$Pu~\cite{JLerendegui:RRR}. The same methodology was then applied to TOF measurements on other actinides~\cite{Alcayne:23,Alcayne:24}. A hybrid approach, so-called resonance weighting factor (RWF), has been also been recently proposed~\cite{PerezMaroto:Thesis}. However, while in the RRR the cross section is dominated by the s-wave contribution, in the energy range of interest in the URR the $p$- and $d$-wave neutrons also contribute to the cross section (see Sec.~\ref{sec:FITACS}). Neutrons with different orbital momenta $\ell$ and higher $E_{n}$ populate resonances with different spin and parity and as a consequence the decay pattern may change significantly.

The impact of different cascade patterns in the capture detection efficiency ($\varepsilon_c$) of C$_6$D$_6$ detectors has been quantitatively evaluated in a recent systematic study for a large range of nuclei~\cite{Mendoza:23}. For the case of $^{242}$Pu, the standard deviation in the unweighted detection efficiency was found to be negligible (0.7\%) between $s$- and $p$-wave resonances for $E_{n}$= 1~keV and $E_{thr}$=150~keV. Nevertheless, in the present work we study the neutron energy range up to $E_{n}$= 600~keV and, as described in Sec.~\ref{sec:FITACS}, $d$-wave resonances are also expected to contribute at higher $E_n$.

In order to assess the neutron-energy dependent capture efficiency in this work on the basis of simulations, we have computed the detection efficiency for capture cascades originating in resonances with all allowed spin and parities, populated by neutrons in the energy range of interest ($E_n$=1--600~keV). The accuracy of these simulations was validated against the experimental response in our previous work on the RRR (See Fig. 4 of Ref.~\cite{JLerendegui:RRR}). The impact of the detection threshold in $\gamma$-ray deposited energy $E_{thr}$ has also been explored since the final analysis of the data has been performed using  different values of $E_{thr}$ depending on the neutron energy range. As explained later in Sec.~\ref{sec:EnergyLimit}, the low neutron energy range $E_n$=1-100~keV was analyzed using  $E_{thr}$=150~keV, the same one used for the analysis of the RRR~\cite{JLerendegui:RRR}, to ensure an accurate normalization to this region. In contrast, the high energy range $E_n$=100--600~keV was analyzed with $E_{thr}$=750~keV to suppress and minimize the background due to the inelastic and fission channels, respectively.

 The cascades were simulated using the statistical model code NuDEX~\cite{Mendoza:20,Mendoza:23}, which takes all the information experimentally known -- level scheme, $\gamma$-ray transition probabilities, and internal conversion factors -- from the RIPL-3 database~\cite{RIPL}, which takes them from ENSDF~\cite{ENSDF}. The remaining values are generated randomly according to statistical models based on the Photon Strength Functions and Level Density Parameters of Ref.~\cite{Laplace:2016}. More details on the modelling of $^{242}$Pu(n,$\gamma$) cascades in this work can be found in Refs.~\cite{JLerendegui:RRR,JLerendegui:Thesis}. Using a fixed parametrization of the statistical model parameters, 100 sub-realizations of the same nucleus~\cite{Becvar:1998} were simulated. In particular, decay cascades were generated from resonances with spins and parities $J^{\pi}$= 1/2$^+$, 1/2$^-$, 3/2$^-$, 3/2$^+$, 5/2$^+$ populated by neutrons with angular momenta $\ell=0,1,2$ at energies $E_n$=1, 10, 50, 100, 250 and 500~keV. For each sub-realization, a total of 10$^{5}$ cascades were generated. Then, Geant4 was applied to simulate the response of the C$_6$D$_6$ detection setup to each individual cascade. The obtained spectra have allowed us to quantify the average capture detection efficiency ($\langle \varepsilon_c \rangle$) for each combination of $J^{\pi}$, $E_n$ and $E_{thr}$, as well as the relative standard deviation of the different sub-realizations ($\sigma_r$).

\begin{figure}[!h]
\begin{center}
\includegraphics[width=8.5cm]{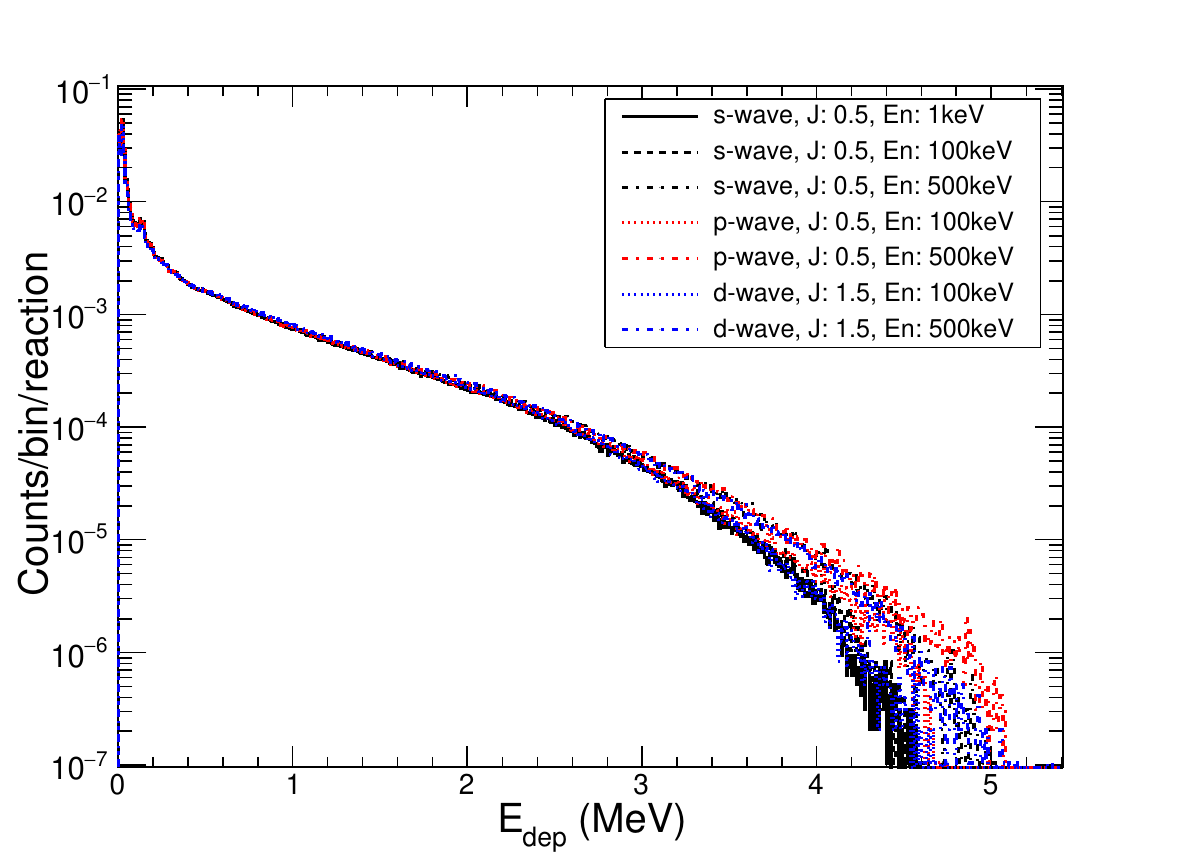}
\includegraphics[width=8.5cm]{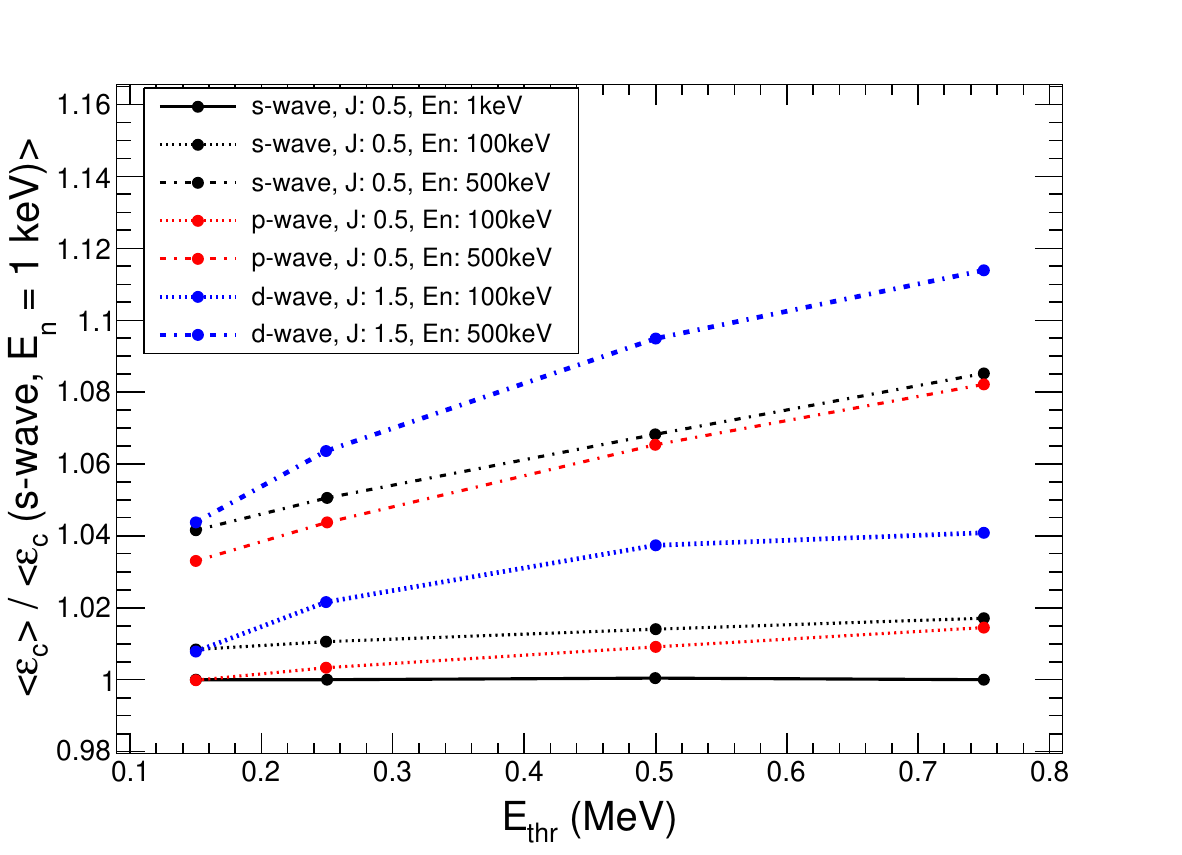}
\caption{Top: Simulated response of the four C$_6$D$_6$ detectors for (n,$\gamma$) resonances with different spin, parity and neutron energy. Bottom: Average efficiency $\langle \varepsilon_c \rangle$ obtained from the simulated cascades for different $J^{\pi}$ and $E_n$ as a function of $E_{thr}$ normalized to the value for low energy s-wave resonances.}
\label{Fig_CascadePartialWaves}
\end{center}
\end{figure}

The average response of the $C_6D_6$ detectors to the simulated cascades is shown in the top panel of Fig.~\ref{Fig_CascadePartialWaves} for selected combinations of $J^\pi$ and $E_n$. This figure compares the response to cascades of low energy ($E_n$ = 1~keV) $s$-wave resonances with respect to cascades produced by high energy neutrons ($E_n$ = 100 and 500~keV) with different angular momenta. Only one $J^{\pi}$ is shown for $\ell=1,2$ for simplicity. Based on these detector responses, the variation of the average efficiency $\langle \varepsilon_c \rangle$ for the different resonance $J^{\pi}$and $E_n$ has been studied as a function of $E_{thr}$. The results, presented in the bottom panel of the same figure, indicate that the neutron energy and the threshold have the largest impact in the variation of the efficiency, leading to a maximum increase of 8--11\% for $E_{n}$=500~keV and $E_{thr}$=750~keV with respect to the efficiency at low neutron energies ($E_{n}$=1~keV).
%As explained before, the average weighting factor method used this work, assumes that $\varepsilon_c$ varies with the neutron energy. The resulting efficiency ratio for $E_{n}$=100 and 500~keV is indicated in the bottom panel of Fig.~\ref{Fig_CascadePartialWaves} with straight orange lines. 

%\begin{table}[!t]
%\begin{center}
%\caption{Efficiency for $^{242}$Pu(n,$\gamma$) events ($\varepsilon_c$) as a function of the neutron energy (E$_n$)  normalized to the efficiency at E$_n\leq$1~keV. The results from the simulated cascades with the final values of $E_{thr}$ are compared to the value assumed in this work. The last column shows the relative deviation.}
%\begin{tabular}{ccccc} 
%\hline \hline
%&    \multicolumn{3}{c}{\textbf{$\varepsilon_c/\varepsilon_c (E_{n}=1~keV)$}}   & \\
%\textbf{E$_n$ (keV)} &  \textbf{150~keV}  &  \textbf{750~keV} & \textbf{This work} &  \textbf{Dev. (\%)}  \\
%\hline
%1        &   1.000                &   -                   &   1.000             &   0     					\\
%10       &   0.998                &   -                   &   1.002             &   0.4   					\\
%50       &   0.998                &   -                   &   1.010             &   1.2   					\\
%100      &   1.006                &   1.027               &  1.020              &  1.4-0.8 				\\
%250      &   -                &   1.054               &  1.050              &   0.4 				\\
%500      &   -                    &   1.105               &   1.099             &   0.6 	   				\\
%\hline \hline  
%\end{tabular} 
%\label{Table_Eff}
%\end{center}
%\end{table}

\begin{table}[!t]
\begin{center}
\caption{Efficiency variation factor $F_{\varepsilon,c}^{thr} (E_n)$ as a function of the neutron energy (E$_n$) for the final values of $E_{thr}$ used in this work (see Sec.~\ref{sec:EnergyLimit}). The theoretical dependence given by TED technique after the application of the conventional PHWT is given in the last column.}
\begin{tabular}{ccccc} 
\hline \hline
&    \multicolumn{2}{c}{\textbf{$F_{\varepsilon,c}^{thr} (E_n)$}}  \\
\textbf{E$_n$ (keV)} &  \textbf{$E_{thr}$=150~keV}  &  \textbf{$E_{thr}$=750~keV} & \textbf{TED} \\
\hline
1        &   1.000                &   -              &  1.000  \\
10       &   0.998                &   -              &  1.002  \\
50       &   0.998                &   -              &  1.010 \\
100      &   1.006                &   1.027          &  1.020 \\
250      &   -                    &   1.054          &  1.050  \\
500      &   -                    &   1.105          &  1.099 \\
\hline \hline  
\end{tabular} 
\label{Table_Eff}
\end{center}
\end{table}

\begin{figure}[!h]
\begin{center}
\includegraphics[width=8.5cm]{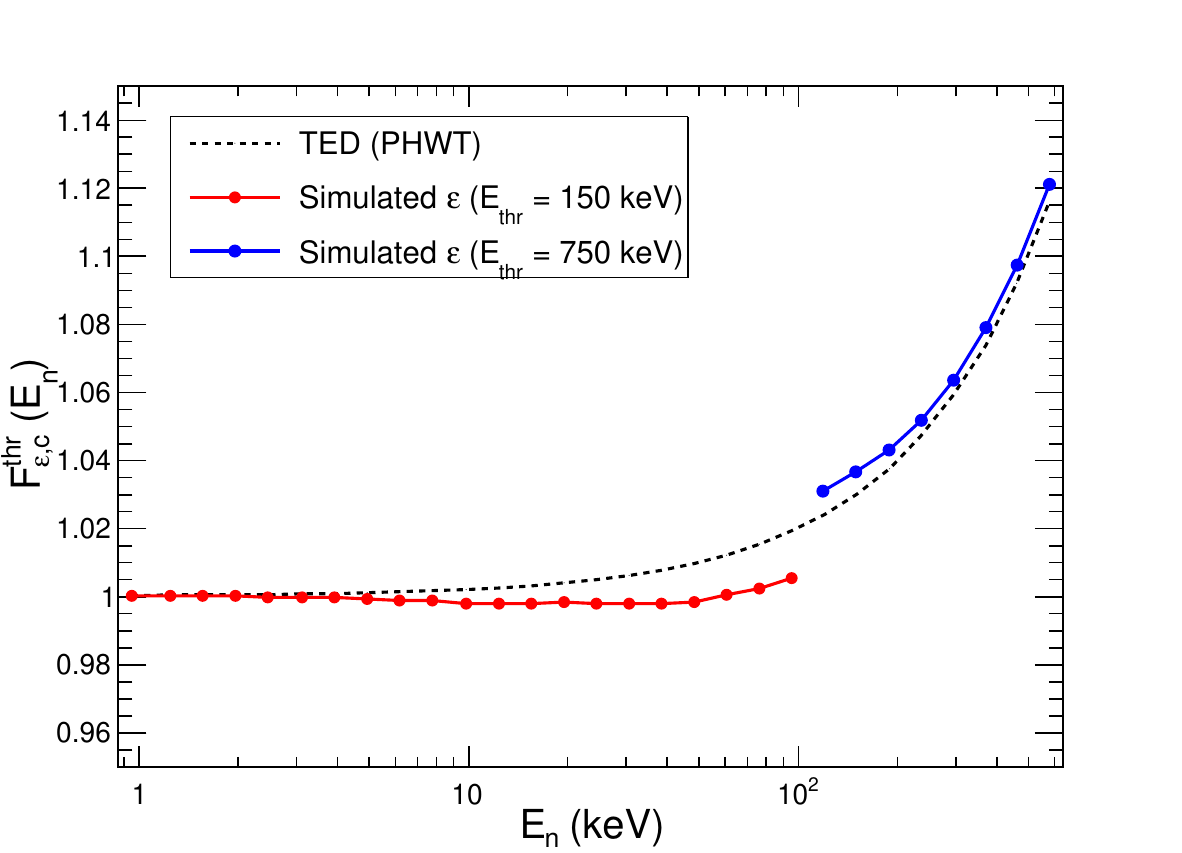}
\caption{Efficiency variation factor $F_{\varepsilon,c}^{thr} (E_n)$ as a function of the neutron energy (E$_n$) interpolated from the results of the simulations (NuDEX + Geant4). The red and blue curves correspond, respectively, to the variation with $E_{thr}$=150~keV and 750~keV. These values have been used to analyze the neutron energy range below and above $E_n$=100~keV, respectively. The dashed line shows, for comparison, the energy dependence given by TED technique after the application of the conventional PHWT.}
\label{Fig_EffRatio_vs_Energy}
\end{center}
\end{figure}

Following the results of the simulations, we have calculated $\varepsilon_c$($E_{n}$) from the dependence of the efficiency with $J^{\pi}$, $E_{n}$ and $E_{thr}$. For this purpose, $\varepsilon_c$($E_{n}$) has been computed by weighting the efficiency for each $J^{\pi}$ by the neutron-energy dependent contribution of each angular momenta $\ell$, discussed in Sec.~\ref{sec:FITACS}. The same values of $E_{thr}$ used for the final analysis in the different energy ranges (see Sec.~\ref{sec:EnergyLimit}), have been applied for the assessment of the efficiency. The resulting neutron-energy dependency of the efficiency, given by $F_{\varepsilon,c}^{thr} (E_n)=\varepsilon_c(E_{n})^{thr}/\varepsilon_{c}^{thr}(E_n=0\  {\rm~keV})$, are shown in Fig.~\ref{Fig_EffRatio_vs_Energy} and Table~\ref{Table_Eff}. The change of efficiency with neutron energy is negligible ($\leq$0.2\%) for E$_{n}<$50~keV. In the range $E_{n}$=50-100~keV, the relative variation of the efficiency stays below 0.6\%. At higher neutron energies, where the data have been analyzed with $E_{thr}$=750~keV, the relative increase of the efficiency is more sizable, increasing in 2.7\% at $E_{n}$=100~keV and up to 10.5\% at 500~keV. It is remarkable that the neutron-energy dependent variation of the efficiency is close to the linear dependency given by the TED technique, $\varepsilon_c(E_{n})=k \cdot (S_n$+E$_n$), after the application of the conventional PHWT.

The final systematic uncertainty associated with the determination of the efficiency have a twofold origin. First, the systematic uncertainty in the determination of the average weighting factor, that allows to assess the efficiency at low neutron energies, is assumed to be 2\% and independent of the neutron energy. The reader is referred to the analysis of the RRR~\cite{JLerendegui:RRR} for a more detailed description on the sources of uncertainty. Second, following the results of the simulation study in this Section, we have added an additional neutron-energy dependent systematic uncertainty associated with the standard deviation in the efficiency resulting from differences between the individual sub-realizations $\sigma_r$. This uncertainty is always below 1\% for $E_{thr}$=150~keV, which applies to neutron energies below 100~keV, and increases up to a maximum value of 1.5\% for $E_{thr}$=750~keV, which affects the neutron energy range E$_n>$100~keV. The uncertainty in the contribution of different neutron orbital momenta as a function neutron energy has also been studied and found to have a negligible impact in the efficiency ($\leq$0.2\%). This result is consistent with the small variations in efficiency observed between different spin and parities for the same neutron energy (see Fig.~\ref{Fig_CascadePartialWaves}).

\subsection{\label{sec:Backgrounds} Determination of the background}

%\item Directly measured: Beamoff fitted + Dummy (highlight that dummy almost = to empty : importance of thin backings) 
%\item Beamoff substracted individually for each sample
%\item Pb(n,$\gamma)$ Ancilliary measurement: Fitted In-beam g-rays and scattered neutron contributions
%\item scaling gammas (Geant4 simulations) and neutron contribution (ratio of elastic cross sections)

\begin{figure}[!b]
\begin{center}
\includegraphics[width=8.3cm]{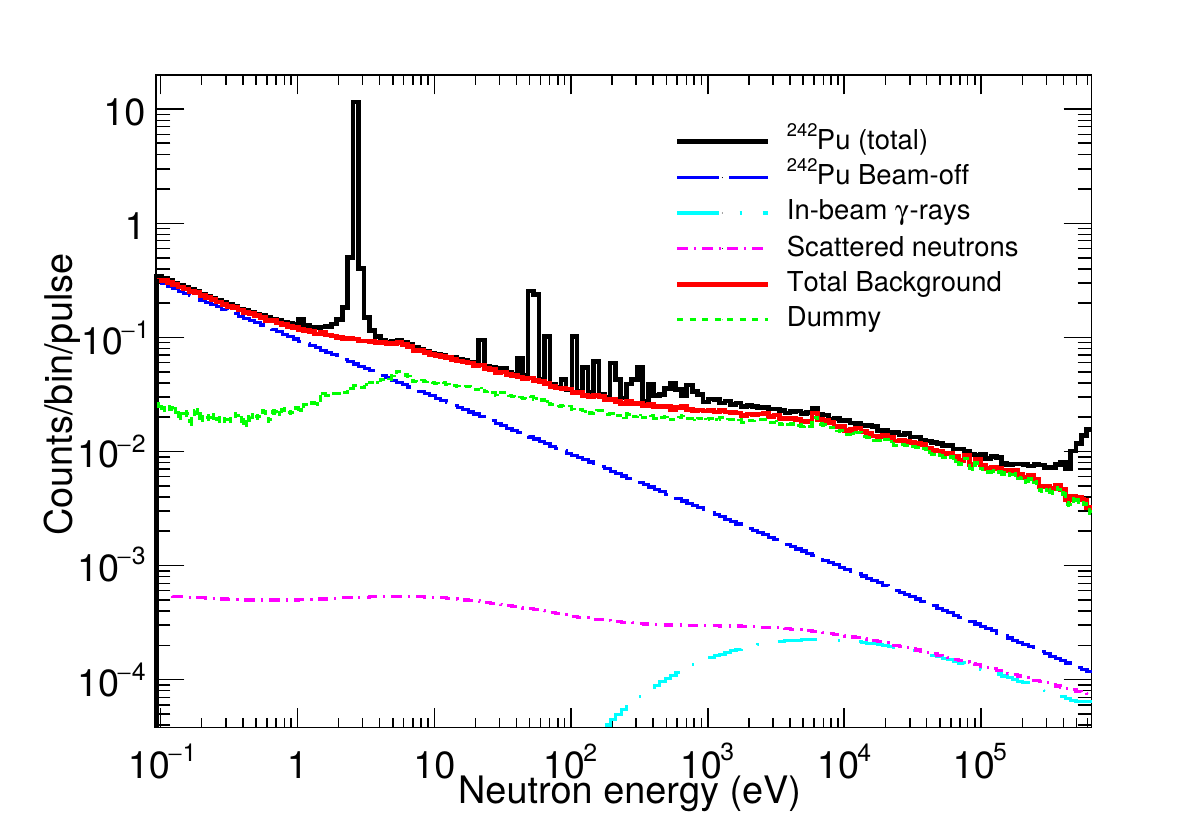}
\includegraphics[width=8.6cm]{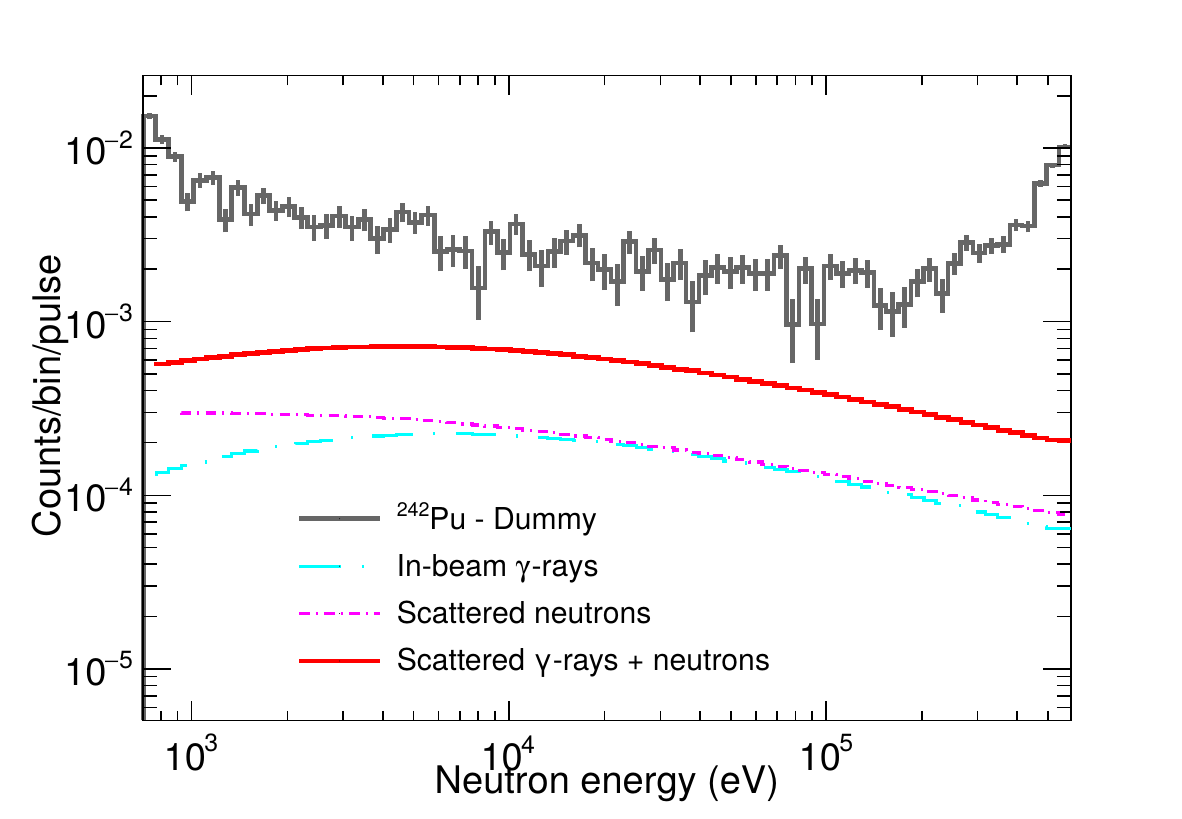}
\caption{Top: Total counting rate of $^{242}$Pu per proton pulse and contribution of the different background components using a detection threshold ($E_{thr}$) of 150~keV. The counting rate in the URR (E$_{n}~>$~1~keV) is dominated by
the beam related background (dummy) (see text for details). The beam-off contribution to the dummy background has been subtracted. Bottom: Zoom in the URR showing the counting rate after the dummy 
is subtracted compared to the remaining background (i.e. scattered in-beam neutrons and gammas).}
\label{Fig_summary_bckgs}
\end{center}
\end{figure}

Assessing the background in our measurement is the most critical point for the analysis of the URR because both background and capture show a smooth 
shape without resonant structures, and the former dominates.
A series of ancillary measurements were carried out to assess the different sources of background, displayed together
with the total spectrum of the $^{242}$Pu measurement in the top panel of Fig.~\ref{Fig_summary_bckgs}. In this figure and hereafter,
unweighted counts are shown unless specified. The total background $B_{T}$ as a function of the neutron energy, shown as a red line in Fig.~\ref{Fig_summary_bckgs},
is given by:
\begin{equation}
 B_{T} (E_{n})  = B_{\rm dummy} + B^{\rm Pu}_{\rm off} + B^{\rm Pu}_{\rm n,n} + B^{Pu}_{\rm \gamma,\gamma}, 
 \label{eq:bckg}
\end{equation}
where $B_{\rm dummy}$ is the dummy background, $B^{Pu}_{\rm off}$ is the beam-off contribution,
and $B^{\rm Pu}_{\rm n,n}$ and $B^{Pu}_{\rm \gamma,\gamma}$ are the background due to neutrons and in-beam $\gamma$-rays scattered 
in the $^{242}$Pu targets.

Among the different contributions, some can be directly assessed from the measurements. First, the beam-off background ($B_{\rm off})$ was estimated from a measurement with no beam and fitted to a constant value as a function of time with a relative uncertainty below 1\%. This background component was first normalized to the number of neutron bunches in each measurement ($^{242}$Pu, $^{197}$Au, Pb, dummy, empty) and then subtracted. In addition, the measurement of the dummy target includes the beam-related background ($B_{\rm dummy})$ in Eq.~(\ref{eq:bckg}) accounting for neutrons 
and $\gamma$-rays scattered in the beam line, vacuum windows and target backings. Thanks to the use of thin backings, the increase in background in the URR  with respect to the situation with no target in the beam (\textit{empty} background in the following) is just 5\%. This implies that a conservative estimate of deviation
of 10\% in the Al backing mass of the dummy sample with respect to the ones in the $^{242}$Pu target would affect the overall background by just 0.5\%.
This is crucial in the URR, where the dummy background accounts for 75-80\% of the total measured counts with the $^{242}$Pu target (see Table~\ref{Table_ContributionsURR}).

Besides the directly measured backgrounds $B_{\rm dummy}$ and $B^{\rm Pu}_{\rm off}$, two additional contributions from Eq. (\ref{eq:bckg}) are shown in the top panel of Fig.~\ref{Fig_summary_bckgs}, 
related to the neutron  $B^{\rm Pu}_{\rm n,n}$ and in-beam $\gamma$-rays $B^{\rm Pu}_{\rm \gamma,\gamma}$ scattered 
in the $^{242}$Pu targets. These two components can not be measured directly but have been inferred from the 
measurement of a Pb target. This target features, similarly to $^{242}$Pu,
a high $\gamma$-ray interaction probability due to its large atomic number Z and high density but presents a negligible capture to elastic neutron 
cross section ratio. From the measured counting rate with the Pb target, the contributions of neutrons and $\gamma$-rays were separately scaled to the $^{242}$Pu sample, as explained in the following, 
because their relative contribution changes between the Pb and the $^{242}$Pu samples.

\begin{figure}[!t]
\begin{center}
\includegraphics[width=8.55cm]{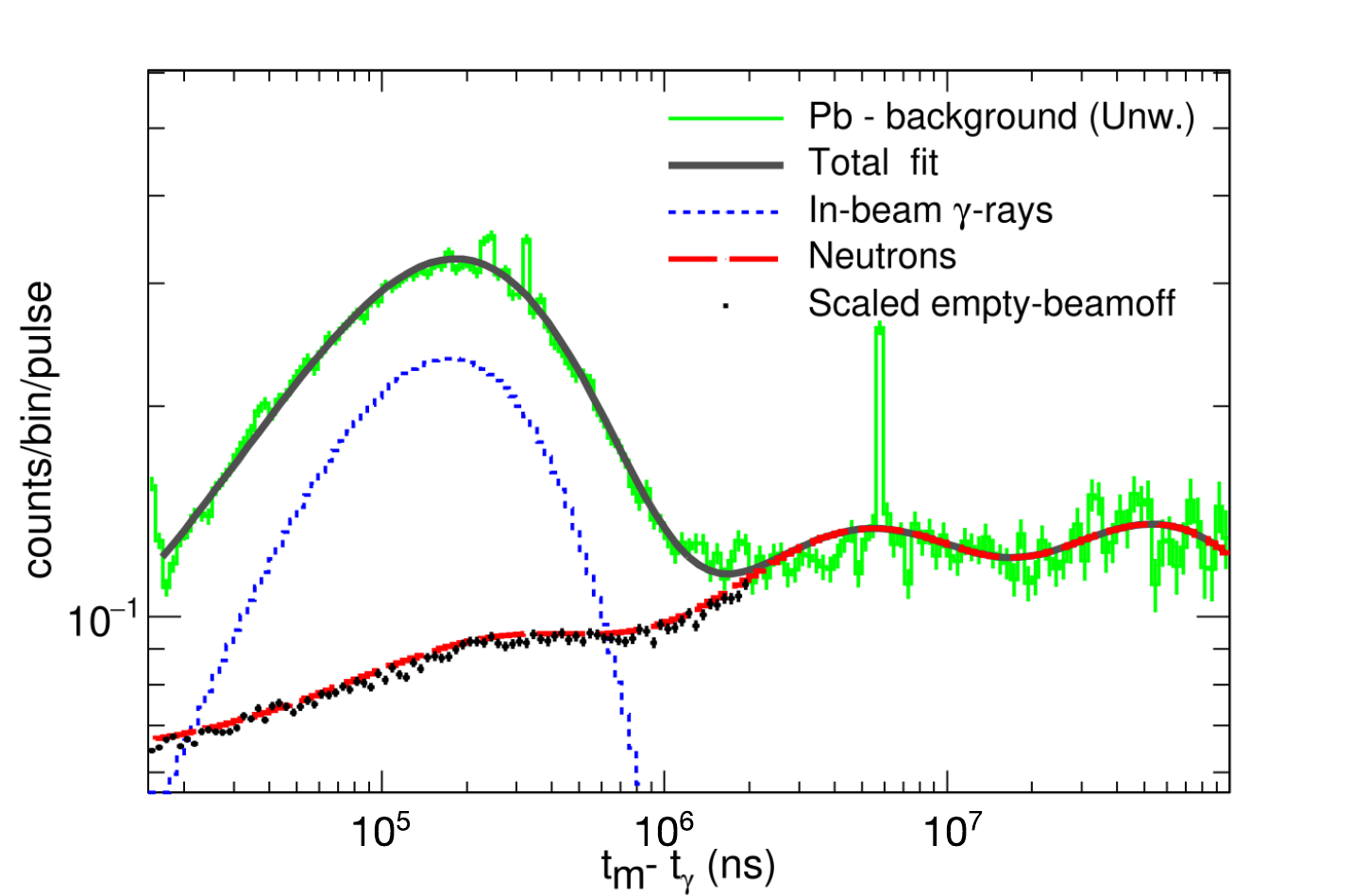}

\includegraphics[width=8.3cm]{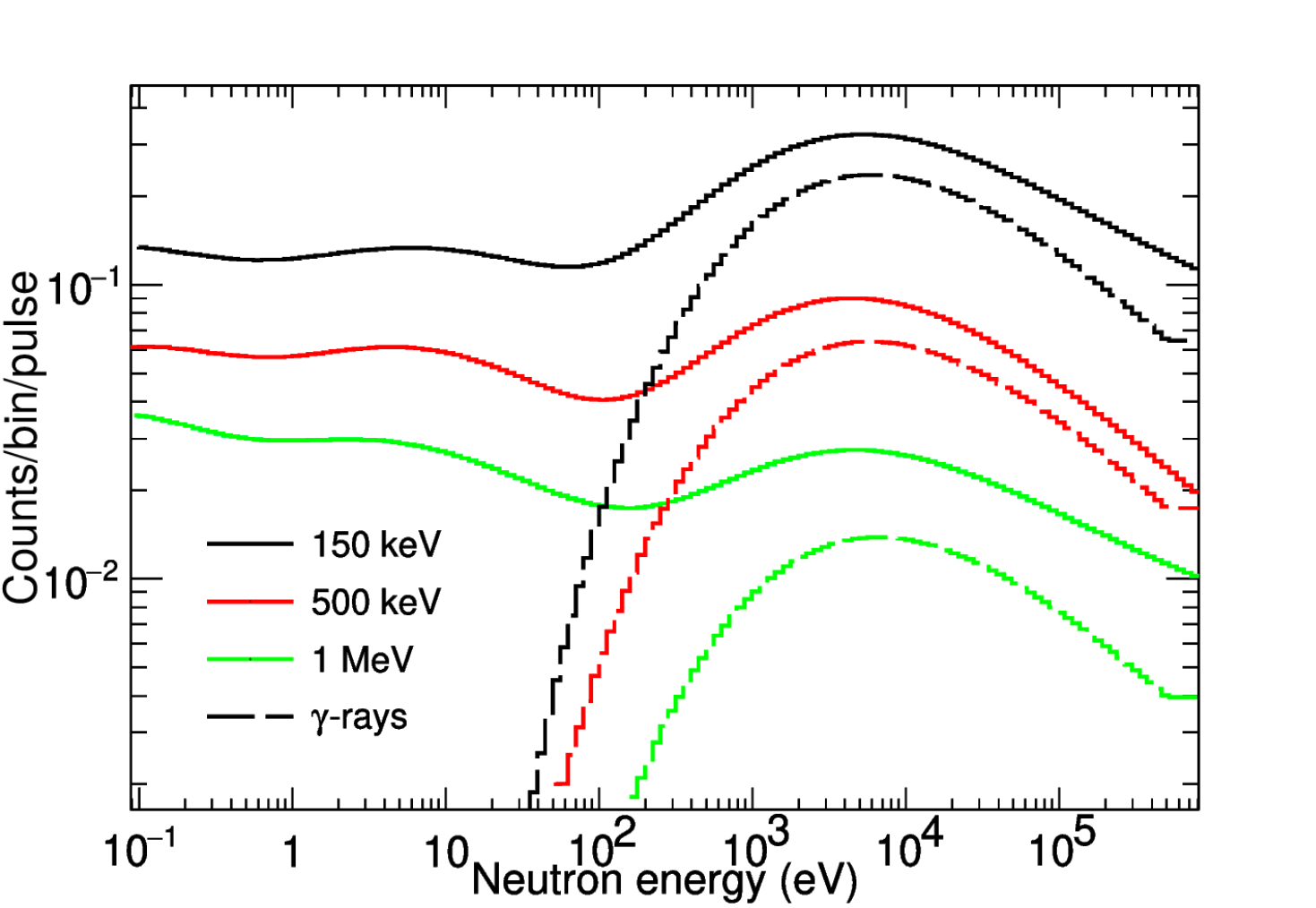}
\caption{Top: Pb counting rate as a function of the time-of-flight ($E_{thr}$=150~keV) fitted to the expression in Eq.~(\ref{eq:FitPb}) (grey curve). 
The contribution of the in-beam $\gamma$-rays (blue) and the remaining scattered neutron component (red)
have been fitted separately. The shape of the scattered neutron background at $t_m-t_\gamma<$1.5~ms is extracted from a fit of the measured counting rate with an empty holder.
Bottom: Fitted total Pb counting rates (solid) and in-beam $\gamma$-ray contribution (dashed) as a function of neutron energy for different detection thresholds.}
\label{Fig_t2e_FitPb}
\end{center}
\end{figure}

The total counting rate measured with the lead target as a function of the time-of-flight was separated in two contributions 
$C^{Pb}(t)~=C^{Pb}_{\gamma,\gamma}(t) + C^{Pb}_{n,n}(t)$. First, in order to remove the fluctuations and capture resonances, the total counting rate $C^{Pb}(t)$ was fitted to the following empirical formula: 
 \begin{equation}
 C^{Pb}(t)~=~k_{0}~+~\sum_{i=1}^{3} a_{i}\cdot(1-e^{-b_{i}t})\cdot e^{-c_{i}t},
 \label{eq:FitPb}
 \end{equation}
For time $t_m-t_\gamma>$1.5~ms (corresponding to $E_n\lesssim$100 eV), the counts in the Pb measurement are only due to scattered neutrons $C^{Pb}(t)=C^{Pb}_{n,n}(t)$ because simulations~\cite{Guerrero:2013,LoMeo:2015} show that all the $\gamma$-rays
arrive before that time. At shorter time-of-flights, assuming that the empty background is not sensitive to in-beam $\gamma$-rays due to the absence of heavy elements, the neutron component $C^{Pb}_{n,n}(t)$ is 
obtained from the scaling of the empty spectrum to match the Pb measurement for $t_m-t_\gamma>$1.5~ms (see Fig.~\ref{Fig_t2e_FitPb}). This scaling is justified as the dependence of the counting rate $t_m-t_\gamma>1.5$~ms is very similar in both Pb and empty spectra.
Then, the difference between lead and the scaled empty measurement for larger $E_n$ gives the contribution from the $\gamma$-rays $C^{Pb}_{\gamma,\gamma}(t)$. 

Following this approach, the fit of the total Pb counting rate from Eq.~(\ref{eq:FitPb}) is shown in grey in the top panel of Fig.~\ref{Fig_t2e_FitPb}. In the same plot, the extracted contributions of
scattered neutron $C^{Pb}_{n,n}$ and $\gamma$-rays $C^{Pb}_{\gamma,\gamma}$ are shown as red and blue lines, respectively. 
The lower panel of Fig.~\ref{Fig_t2e_FitPb} shows how the contribution of scattered neutrons
and $\gamma$-rays changes in the measured Pb data with the detection threshold ($E_{thr}$) applied, the $\gamma$-ray background being less relevant as $E_{thr}$ increases. This
confirms that the fitted Pb counting rates are actually a sum of two different contributions. 

%The fitted counting rate $C^{Pb}(t)$ accounts for two different background sources $C^{Pb}(t)~=C^{Pb}_{\gamma}(t) + C^{Pb}_{n}(t)$. From this fit, we extracted $ C^{Pb}_{\gamma}(t)$, the contribution of
%scattered in-beam $\gamma$-rays, indicated with a blue line in the top panel Fig.~\ref{Fig_t2e_FitPb} is known from simulations to arrive to EAR1 at $t\lesssim$1.5~ms 
%(i.e. neutron energies above 100~eV)~\cite{Guerrero:2013, LoMeo:2015}. 
%The contribution of scattered neutrons $C^{Pb}_{n}(t)$ (red line in the top panel of Fig.~\ref{Fig_t2e_FitPb}) was obtained with the following approach. For times $t>$1.5~ms,
%when no in-beam $\gamma$-rays are present, it is directly the fitted t Pb(n,$\gamma)$ counting rate. At shorter time-of-flights (i.e. below the $\gamma$-ray bump), the background due to scattered 
%neutrons should not vanish but also contribute to the measured Pb, according to simulations of EAR1 performed with Geant4~\cite{Zugec:Background}. Its shape in this time interval was
%fitted from the empty background and scaled to match the Pb(n,$\gamma)$ counting rate at t=2~ms, where the black and red curves merge in Fig.~\ref{Fig_t2e_FitPb}. 

%The smooth dependence of the neutro with the time-of-flight 
%according to simulation of the facility~\cite{Zugec:Background}. 
\begin{figure}[!b]
\begin{center}
\includegraphics[width=8.5cm]{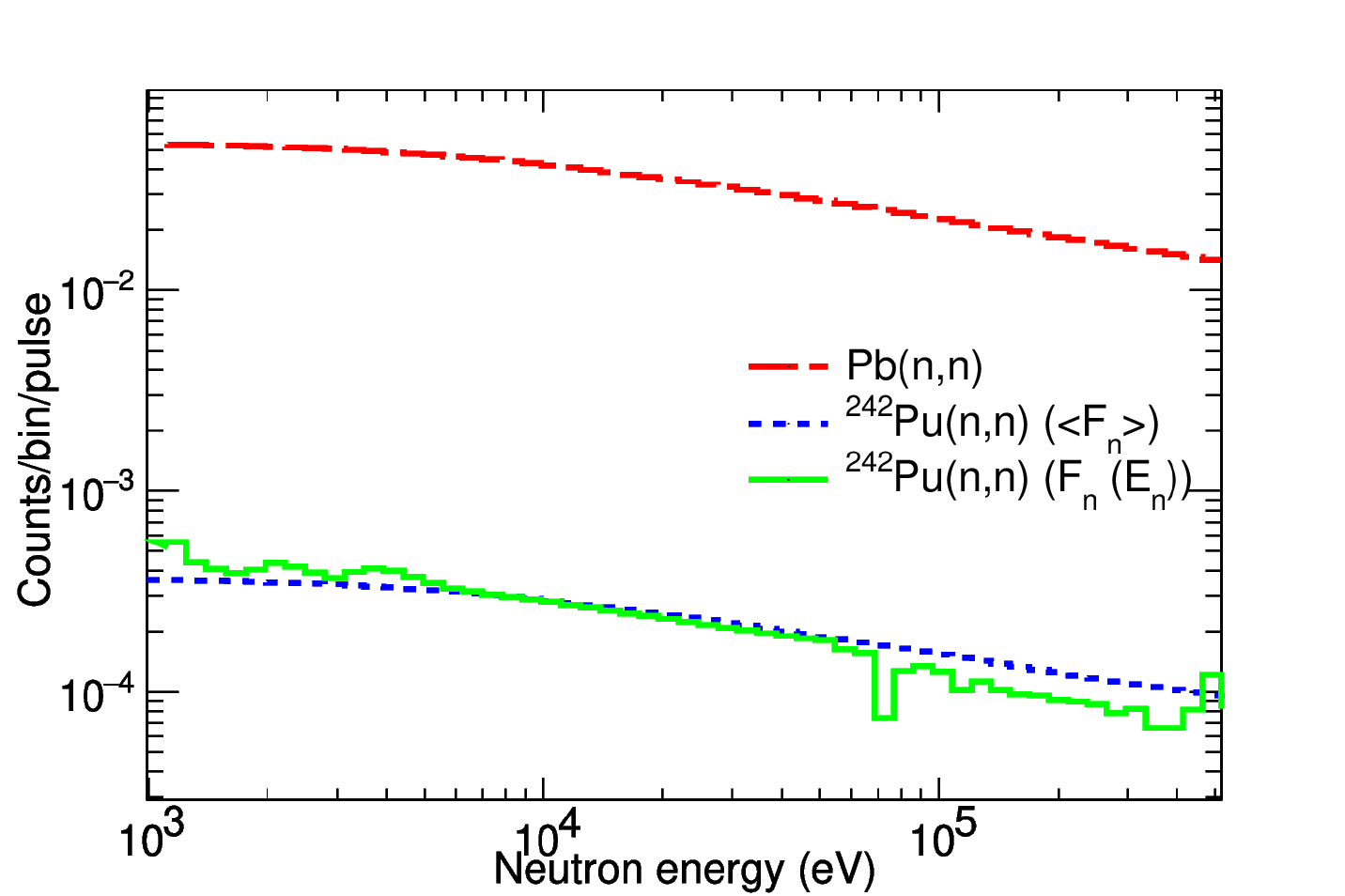}
\caption{Neutron scattering background fitted from the Pb measurement (red dashed curve) scaled to the $^{242}$Pu using the energy dependent scaling factor $F_{n} (E_{n})$ 
(green solid line) and the average factor $\langle F_{n}\rangle$ (blue dotted line). The latter has been taken as the final neutron scattering background in this analysis, shown in Fig.~\ref{Fig_summary_bckgs}.}
\label{Fig_ScalingNeutrons}
\end{center}
\end{figure}
%The time-of-flight distributions $C^{Pb}_{\gamma}(t)$ and $C^{Pb}_{n}(t)$ extracted from the lead measurement were calibrated in neutron energy 
%following the method described in
% in Sec.~\ref{sec:Calibrations}.
The background contributions in our measurement related to the scattering of neutrons $B_{\rm n,n}^{\rm Pu}(E_{n})$ and 
$\gamma$-rays $B_{\rm \gamma,\gamma}^{\rm Pu}(E_{n})$ in the $^{242}$Pu target are then calculated as
\begin{equation}
B_{\rm n,n}^{\rm Pu}(E_{n})~=~F_{n}~\cdot~C_{\rm n,n}^{\rm Pb}(E_{n}), 
\end{equation}
\begin{equation}
B_{\rm \gamma,\gamma}^{\rm Pu}(E_{n})~=~ F_{\gamma}~\cdot~C_{\rm \gamma,\gamma}^{\rm Pb}(E_{n}),
\end{equation}
where $F_{n}$ and $F_{\gamma}$ are the factors needed to scale the scattering yields of neutrons and $\gamma$-rays from Pb to $^{242}$Pu.
 
$F_{n}$ can be extracted from the ratio of neutron elastic cross sections $\sigma^{X}_{n}$ as 
\begin{equation}
 F_{n}(E_{n})=\frac{\sigma^{Pu}_{n}(E_{n})\cdot n_{Pu}}{\sigma^{Pb}_{n}(E_{n})\cdot n_{Pb}}, 
\end{equation}
where $n_{X}$ is the areal density of the samples.
However, simulations of the neutron transport showed that its time-of-flight distribution is smoothened by the moderation of scattered neutrons around 
EAR1 and does not follow directly the neutron energy dependence of scattered neutrons~\cite{Zugec:Background}. Hence it is more realistic to assume an average scaling factor independent of the neutron energy given by 
\begin{equation}
\langle F_{n} \rangle =\big \langle \frac{\sigma^{Pu}_{n}}{\sigma^{Pb}_{n}}\big \rangle \frac{n_{Pu}}{n_{Pb}},
\end{equation}
where the averaging is performed in the energy region under study (1-600~keV). The latter method has been adopted for the final analysis.
Both methods for the neutron scattering scaling are compared in Fig.~\ref{Fig_ScalingNeutrons}.
The difference in the resulting $^{242}$Pu capture yield using these two method serves to estimate the associated uncertainty. The relative contribution of neutron scattering in Fig.~\ref{Fig_summary_bckgs} combined with the deviation between the two methods in Fig.~\ref{Fig_ScalingNeutrons}  yields a systematic uncertainty in the final yield of 1.5\% below 10~keV,
1\% between 10 and 50~keV, 2.5\% between 50 and 250~keV and 1.5\% at higher energies.

\begin{figure}[!h]
\begin{center}
\includegraphics[width=8.52cm]{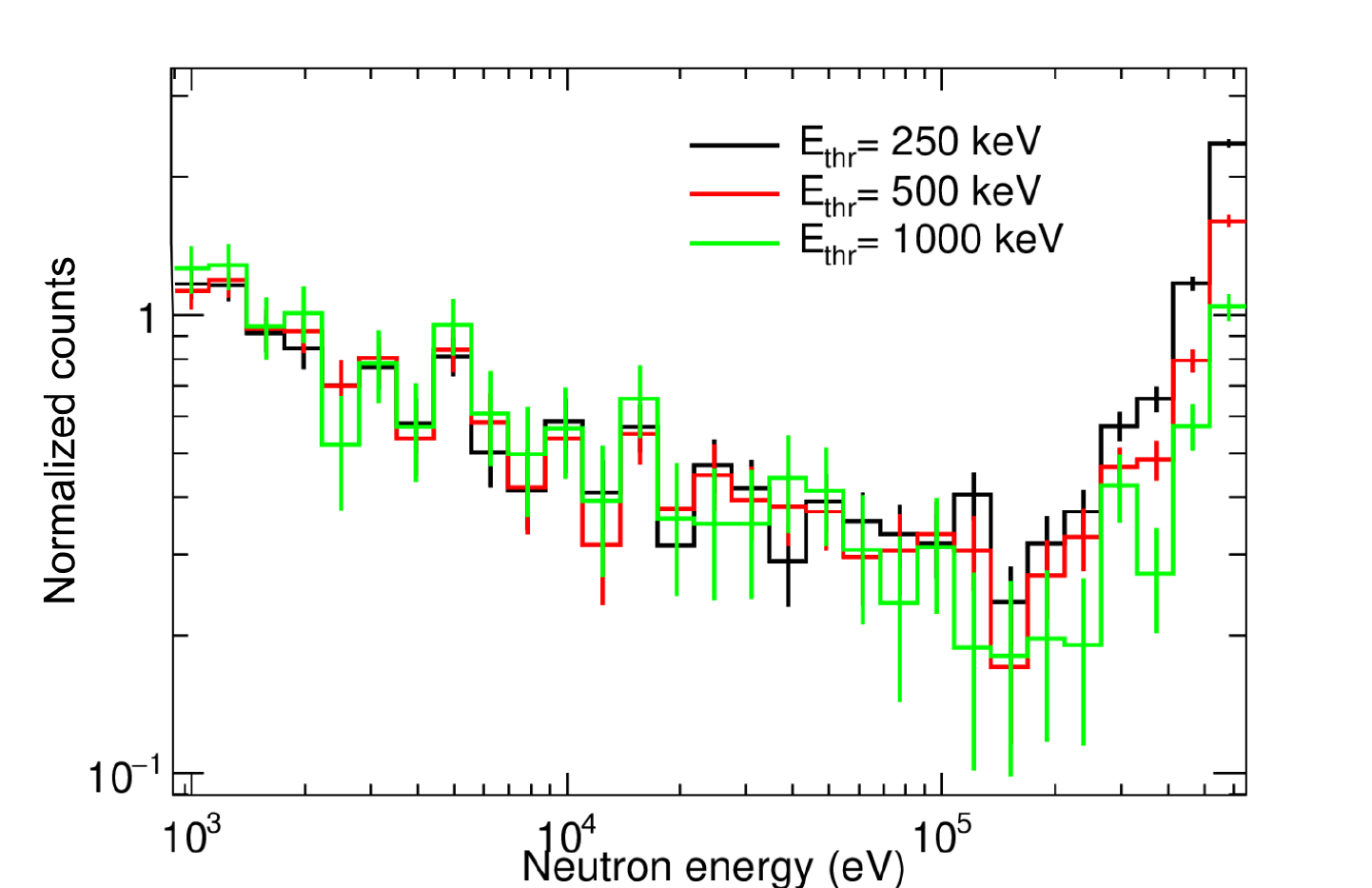}
\includegraphics[width=8.5cm]{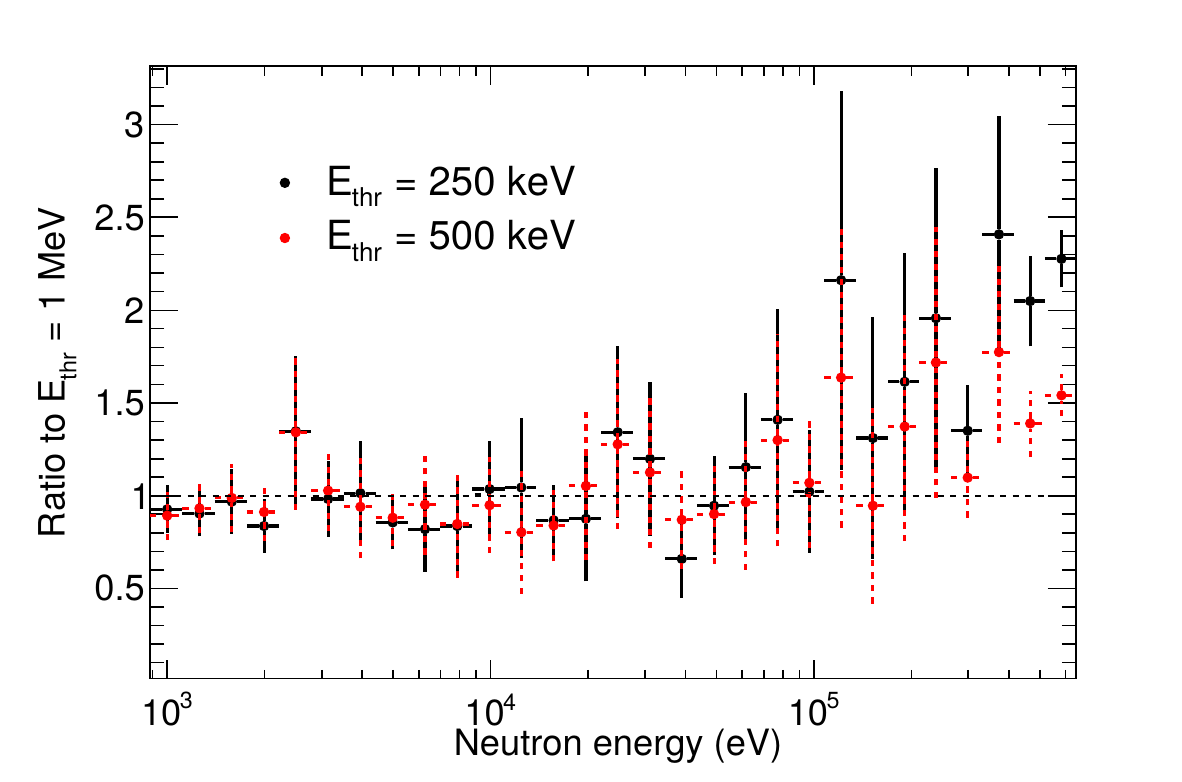}
\caption{Top: $^{242}$Pu($n$,$\gamma$) counting rate per pulse in the URR as a function of the neutron energy obtained with
different thresholds ($E_{thr}$) normalized to the 2.67 eV resonance. The good consistency of the result up to a neutron energy of 100~keV validates the absolute scaling factor of the in-beam $\gamma$-ray background.
The bottom panel shows the ratio of counting rates with respect to the results with $E_{thr}$=1 MeV (the average ratios are summarized in Table~\ref{Table_RatiosAfterScaling}).}
\label{Fig_ValidScalingGammas}
\end{center}
\end{figure}

For the case of the $\gamma$-ray background, the scaling factor $F_{\gamma}$ does not depend on the neutron energy, but only on the detection threshold.
In previous works at n\_TOF-EAR1 $F_{\gamma}$ 
was fitted with the help of black resonance filters (see for instance Refs.~\cite{Lederer:2011,Migrone:2017}). 
However, during the $^{242}$Pu campaign no measurements were performed with filters and $F_{\gamma}$ was instead calculated as
\begin{equation}
F_{\gamma} = F^{abs}_{\gamma}\cdot F^{thr}_{\gamma},
\end{equation}
where $F^{thr}_{\gamma}$ is a detection threshold dependent factor which was calculated using Monte Carlo simulations of the in-beam $\gamma$-rays~\cite{LoMeo:2015} scattered in the Pb, Au and Pu targets
using the Geant4 toolkit~\cite{Geant4_1,Geant4_2}. 
%The simulated energy distribution of in-beam $\gamma$-rays was taken from the recent Geant4 simulations of 
%the n\_TOF-EAR1 facility~\cite{LoMeo:2015}.
On the other hand, $F^{abs}_{\gamma}$ was determined as the absolute scaling factor that leads to consistent results for the
background-subtracted $^{242}$Pu($n$,$\gamma$) counting rate in the URR regardless of:
\begin{enumerate}
 \item the detection threshold set for the analysis and
 \item the use of unweighted or weighted (PHWT) data sets in the background subtraction.
\end{enumerate}
Since these two conditions are very sensitive to the different $\gamma$-ray energy distributions of the in-beam $\gamma$-ray background 
compared to that of the capture cascade, they serve to find the absolute scaling of the in-beam $\gamma$-ray background.
The resulting values for $F_{\gamma}$ are listed in Table~\ref{Table_Scaling}. 
With these values a nice agreement of the background-subtracted counting rate is obtained
up to $E_{n}$=100~keV for different thresholds, as shown in Fig.~\ref{Fig_ValidScalingGammas} and quantified in Table~\ref{Table_RatiosAfterScaling}.  As a result of the values in Table~\ref{Table_RatiosAfterScaling}, a 3\% systematic uncertainty in the cross section was associated with the in-beam $\gamma$-ray background. The data above $E_{n}$=100~keV are not considered in the evaluation of the neutron and in-beam gamma-ray background subtraction since this energy region is also affected by the inelastic and fission backgrounds, which have been assessed and subtracted as explained in Sec.~\ref{sec:EnergyLimit}.

\begin{table}[]
\begin{center}
\caption{Scaling factors for scattered in-beam $\gamma$-rays ($F_{\gamma}$) and neutron ($F_{n}$=$\langle F_{n} \rangle$) backgrounds fitted from the Pb ancillary measurement. The uncertainties in $F_{\gamma}$ are statistical (MC simulations) while the uncertainty in $F_{n}$ is dominated by that of the $^{242}$Pu mass.}
\begin{tabular}{ccccc} 
\hline \hline
 \textbf{$E_{thr}$ (keV) } & \textbf{$F_{\gamma}$ ($\times10^{3}$)}  & \textbf{$F_{n}$ ($\times10^{3}$)}     \\
\hline
250                      &         1.67(7)            &      \multirow{3}{*}{6.8(3)}                \\
500                      &         1.89(8)            &                       \\  
1000                     &         1.98(10)            &                          \\
\hline \hline    
\end{tabular} 
\label{Table_Scaling}
\end{center}
\end{table}

%\begin{figure}[!h]
%\begin{center}
%\epsfig{file=242Pu_sample_assembly_modified_7targets_removed_smallLetters.png, width=5.9cm}
%\includegraphics[width=8.5cm]{Fig5_Au_Pu_Pb_0_1mm}
%\caption{Energy deposition of the in-beam $\gamma$-rays backscattered in different samples of the same thickness to study the threshold dependence
%of the scaling factor of the direclty measured (Pb) to the samples of interest ($^{197}$Au and $^{242}$Pu.}
%\label{Fig_SpectraGammaSamples}
%\end{center}
%\end{figure}

\begin{table}[!t]
\begin{center}
\caption{Average ratios of $^{242}$Pu(n,$\gamma)$ counting rate in the URR ($E_n$=1--100~keV) with the adjusted in-beam $\gamma$-ray background scaling factor~$F_\gamma^{abs}$. Second column: ratio of unweighted counting
rates using different thresholds $E_{thr}$ with respect to the $E_{thr}$= 1000~keV.  Third column: Ratio between weighted and unweighted scaled counts as a function of the detection threshold.
The statistical uncertainty of each ratio is indicated in brackets.}
\begin{tabular}{ccccc} 
\hline \hline
                           & \multicolumn{2}{c}{\textbf{Average ratio ($E_n$=1-100~keV))}}\\
 \hline
 \textbf{$E_{thr}$ (keV) } & \textbf{$E_{thr}$/1 MeV}  & \textbf{Unweighted/Weighted}     \\
\hline
250                      &           1.01(5)               &       0.99(6)                     \\
500                      &           0.99(5)               &       1.00(6)                     \\  
1000                     &           1                     &       1.05(7)                     \\
\hline \hline    
\end{tabular} 
\label{Table_RatiosAfterScaling}
\end{center}
\end{table}

The scaled scattered neutron and in-beam $\gamma$-ray backgrounds represent a relevant fraction of the counting rate remaining after dummy and beam-off contributions have been subtracted
(see the bottom panel of Fig.~\ref{Fig_summary_bckgs}). The relative contribution of $^{242}$Pu($n$,$\gamma$)
and the backgrounds to the total counting rate in the URR are summarized in Table~\ref{Table_ContributionsURR}.

\begin{table*}[]
\begin{center}
\caption{Relative contribution of capture and the different background components in two energy regions in the URR. The result for the lowest and highest energy threshold tested in 
this work are compared. The uncertainties in brackets reflect the systematic uncertainties discussed in the text. Statistical uncertainties are negligible due to the large energy ranges.}
\begin{tabular}{ccccc} 
\hline \hline
                           & \multicolumn{2}{c}{\textbf{Contribution (1--10~keV)(\%)}}  & \multicolumn{2}{c}{\textbf{Contribution (10--100~keV)(\%)}}\\
\hline
\textbf{$E_{thr}$ (keV) } & \textbf{$E_{thr}$=150~keV}  & \textbf{$E_{thr}$=1 MeV}           & \textbf{$E_{thr}$=150~keV}  & \textbf{$E_{thr}$=1 MeV}           \\
\hline
Dummy                      &           75.8(8)               &              80.7(8)       		&           80.1(8)                 &     84.3(8)                               \\
Beam-off                   &           7.50(7)              &               4.45(4)        		&           4.13(4)                 &      2.449(24)                               \\
In-beam $\gamma$-rays      &           0.9(4)               &              0.21(14)             &         1.33(4)                 &       0.31(12)                           \\  
Scattered Neutrons         &           1.13(22)             &             0.79(21)         	&          1.3(3)                 &       1.06(24)                               \\
Capture                    &           14.7(11)                &            13.9(10)          &         13.2(11)                   &       11.9(10)                            \\         \hline \hline    
\end{tabular} 
\label{Table_ContributionsURR}
\end{center}
\end{table*}
\subsection{\label{sec:EnergyLimit} High neutron energy limit}

\begin{figure}[!h]
\begin{center}
\includegraphics[width=8.5cm]{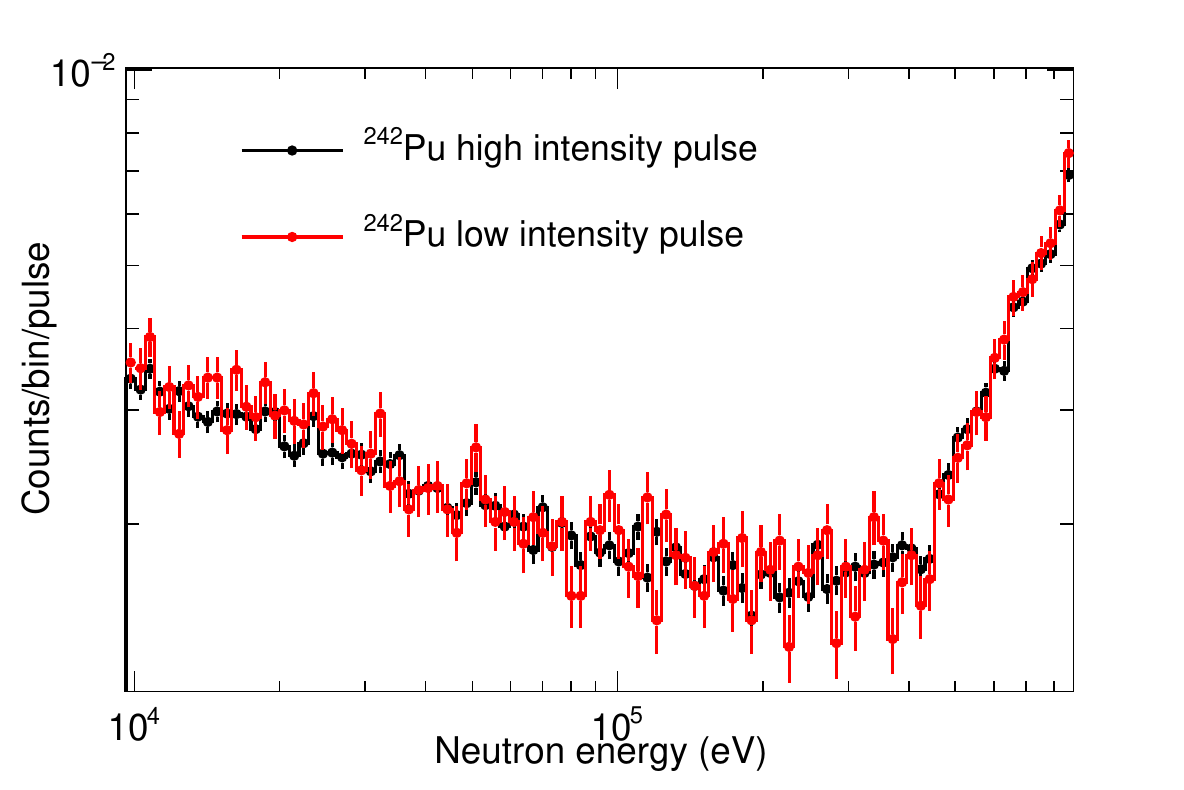}
\caption{Total counting rate as a function of the neutron energy normalized to the nominal number of protons per pulse ($\sim 7 \times 10^{12}$ protons) showing the good agreement of the high-intensity and low-intensity pulses.}
\label{Fig_ParasDedicated}
\end{center}
\end{figure}

Measurements at n\_TOF-EAR1 using the fast C$_{6}$D$_{6}$ detectors allow in principle extracting data up to a neutron energy of about 1~MeV (see for instance Ref.~\cite{Gunsing:2006}). However, as one approaches this maximum energy, the effect of the $\gamma$-flash and the increasing dead time losses
due to the higher counting rates can affect the detector behavior.
Low intensity pulses, featuring about one third of the nominal proton intensity and also a lower $\gamma$-flash have
been used to validate the recorded counting rate at neutron energies above $E_n$=10~keV. The good agreement between the counting rate normalized to number  of protons for the two kind of pulses, shown in Fig.~\ref{Fig_ParasDedicated}, indicates that we are not affected by these issues up to 900~keV. The use of thin backings combined to the fast response of the detectors are key to achieving this result. 

Another limitation at high neutron energies is related to the production of $\gamma$-rays in the sample by neutron induced reactions other than capture, specifically inelastic scattering ($n$,$n'$) and fission ($n$,$f$). The first
$(n, n')$ channel on $^{242}$Pu opens at 45~keV and the contribution of the inelastic scattering increases with the neutron energy~\cite{ENDF}. However, only low-energy $\gamma$ rays are produced in this process and they are eliminated by use of sufficiently high detection threshold in $\gamma$-ray energy.

\begin{figure}[!h]
\begin{center}
\includegraphics[width=8.5cm]{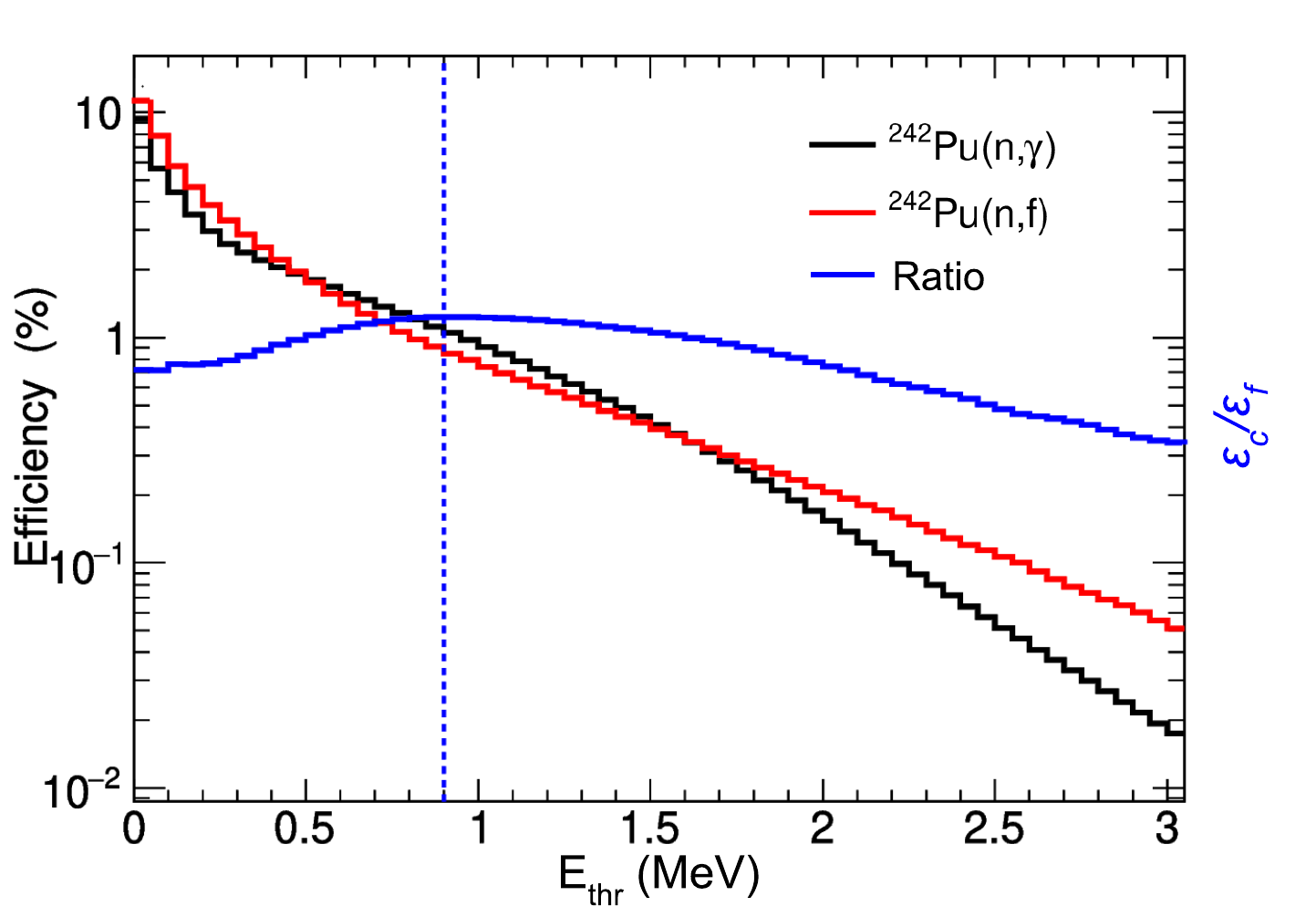}
\includegraphics[width=8.5cm]{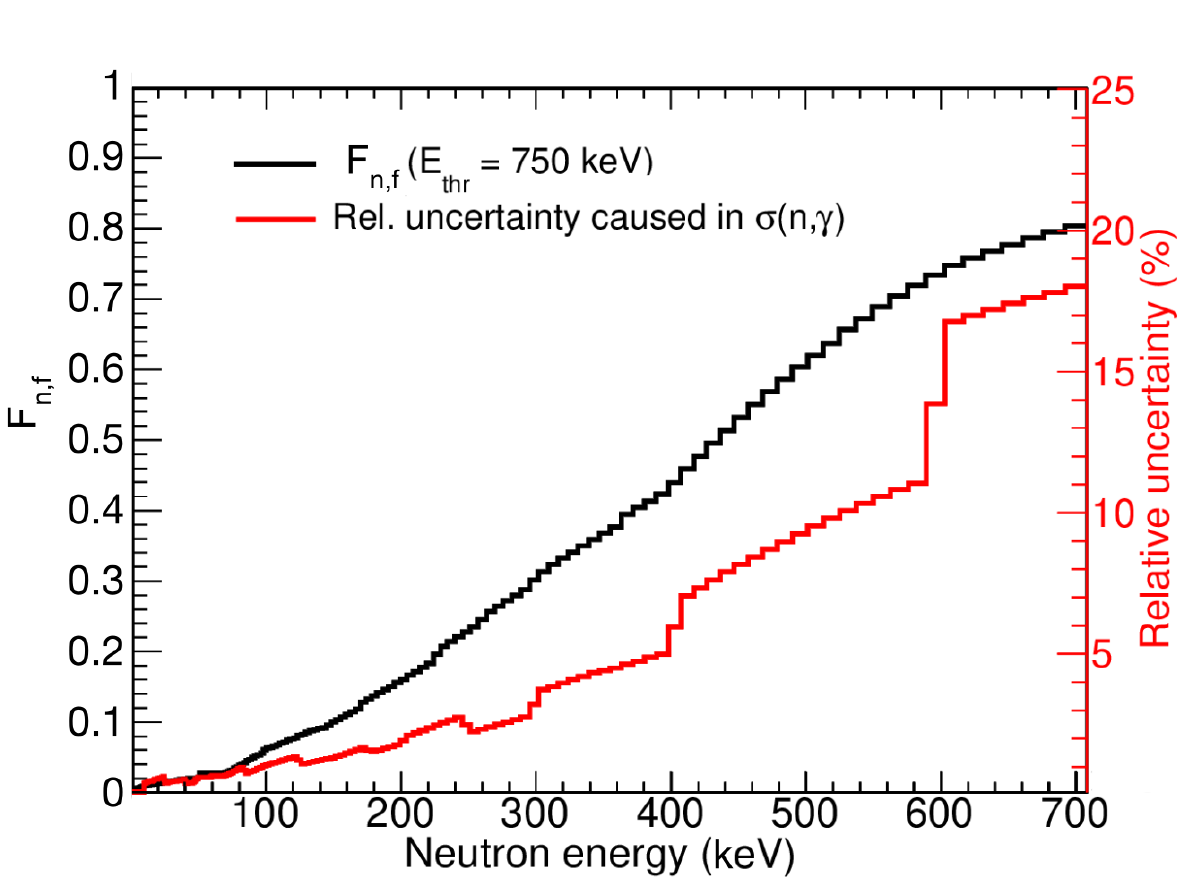}

\caption{Top: Simulated efficiency of one C$_6$D$_6$ detector to capture (black) and fission (red) events on $^{242}$Pu as a function of the deposited energy threshold. The capture-to-fission efficiency ratio $\varepsilon_c/\varepsilon_f$ (blue) is optimized for the threshold indicated with the dashed line. 
Bottom: The black line represents $F_{n,f}$, the relative contribution of the fission channel to the background-subtracted counting rate calculated with a threshold of 750~keV using the ENDF/B-VIII.1 library. The red
curve corresponds to the systematic uncertainty in the cross section associated with the uncertainty in $F_{n,f}$.}
\label{Fig_FissionCaptureSimulations}
\end{center}
\end{figure}

Fission is a more delicate issue since C$_{6}$D$_{6}$ detectors do not allow distinguishing it from capture. 
In order to estimate the contribution from fission in the background-subtracted counting rate -- i.e. after subtracting all the background components defined as $B_T$ in Eq.~(\ref{eq:bckg}) -- we have calculated  the detection efficiency for capture and fission events. The $\gamma$-ray spectra emitted in these two reactions was determined as follows:
\begin{itemize}
 \item Capture: The same capture cascades generated for the correction of the count loss below the detection threshold (see Ref.~\cite{JLerendegui:RRR} for the details) were used
 for this purpose.
  \item Fission: The energy spectrum of fission $\gamma$-rays was obtained using the \textit{GEF} code~\cite{GEF}.
\end{itemize}
The simulated $\gamma$-ray distributions for fission and capture events were coupled to a Geant4 simulation. The resulting efficiency for detection as a function of $E_{thr}$ of the C$_{6}$D$_{6}$ detectors for both fission and capture is shown in the top panel of Fig.~\ref{Fig_FissionCaptureSimulations}. The capture-to-fission efficiency ratio ($\varepsilon_{c}$/$\varepsilon_{f}$)
was also calculated and it is shown in the same figure and for a few values of $E_{thr}$ in Table~\ref{Table_EffFissionCapture}. The efficiency for ($n$,$\gamma$) events is lower than for ($n$,$f$) for detection thresholds below about 500~keV while increases up to 900~keV, for which the maximum $\varepsilon_{c}$/$\varepsilon_{f}$~=~1.23 is reached. For higher $E_{thr}$, 
the detection of fission is favored again relative to capture.

\begin{table}[t]
\begin{center}
\caption{Capture to fission efficiency ratio $\varepsilon_{c}$/$\varepsilon_{f}$ as a function of the detection threshold.
The uncertainties are statistical (from simulations).}
\begin{tabular}{cc} 
\hline \hline 
% &   \multicolumn{3}{c}{\textbf{ $F_{(n,f)}$(\%)}}\\
\textbf{$E_{thr}$ (keV)}    &    \textbf{$\varepsilon_{c}$/$\varepsilon_{f}$}\\ %& \textbf{50-250~keV}  & \textbf{50-250~keV} &  \textbf{250-600~keV}\\
\hline
150                     &    0.758(4)      				      \\%  &         2.2(4)            &       12.0(12)      &   53.0(16)   \\
750                     &    1.210(13)     				      \\%   &         1.4(3)            &       8.5(9)        &   43.0(13) \\
900                     &    1.234(15)     				      \\%   &         1.4(3)           &        8.3(9)        &   42.0(12) \\
1250                    &    1.132(17)                                     \\
%Rich et al.                 && \textbf{8.1}        &\textbf{9.5}       &\textbf{$<$11$^b$}\\ &                  &                         &                      \\
\hline \hline    
\end{tabular} 
\label{Table_EffFissionCapture}
\end{center}
\end{table}

The fraction of the background-subtracted counting rate that comes from ($n$,$f$) events, $F_{n,f}$, was calculated using the simulated efficiency for capture and fission events, as
\begin{equation}
F_{n,f}(E_{n})~=~\frac{\varepsilon_{f} \sigma_{f}(E_{n})}{\varepsilon_{c} \sigma_{\gamma}(E_{n})~+~\varepsilon_{f} \sigma_{f}(E_{n})}, 
 \label{Eq_fis_capt}
\end{equation}
where $\sigma_{f}$ and $\sigma_{\gamma}$ are the evaluated (ENDF/B-VIII.1) fission and capture cross section, respectively. The capture cross section obtained in this work is in good agreement with ENDF/B-VIII.1 between 100--600~keV where the fission correction is relevant, see Sec.~\ref{sec:Comparison}. Thus, no further iteration is needed to correct for the fission contribution. As seen from the bottom panel of Fig.~\ref{Fig_FissionCaptureSimulations}, the contribution of fission is negligible for $E_n<50$~keV and 
becomes larger than capture (F$_{n,f}>$0.5) beyond 450~keV. In the bottom panel of Fig.~\ref{Fig_FissionCaptureSimulations} we also present the systematic uncertainty in the determined (n,$\gamma$) cross section 
associated with the calculated ($n$,$f$) correction $F_{n,f}$ as a function of  neutron energy. This uncertainty has been obtained from the current uncertainties in the $^{242}$Pu(n,f) and $^{242}$Pu($n$,$\gamma$) cross sections according to ENDF/B-VIII.1~\cite{ENDF} 
and becomes the dominant source of systematic uncertainty above 400~keV, restricting the upper energy limit of the present data (see Sec.~\ref{sec:Pucrosssection}). 

\begin{figure}[b]
\begin{center}
\includegraphics[width=8.5cm]{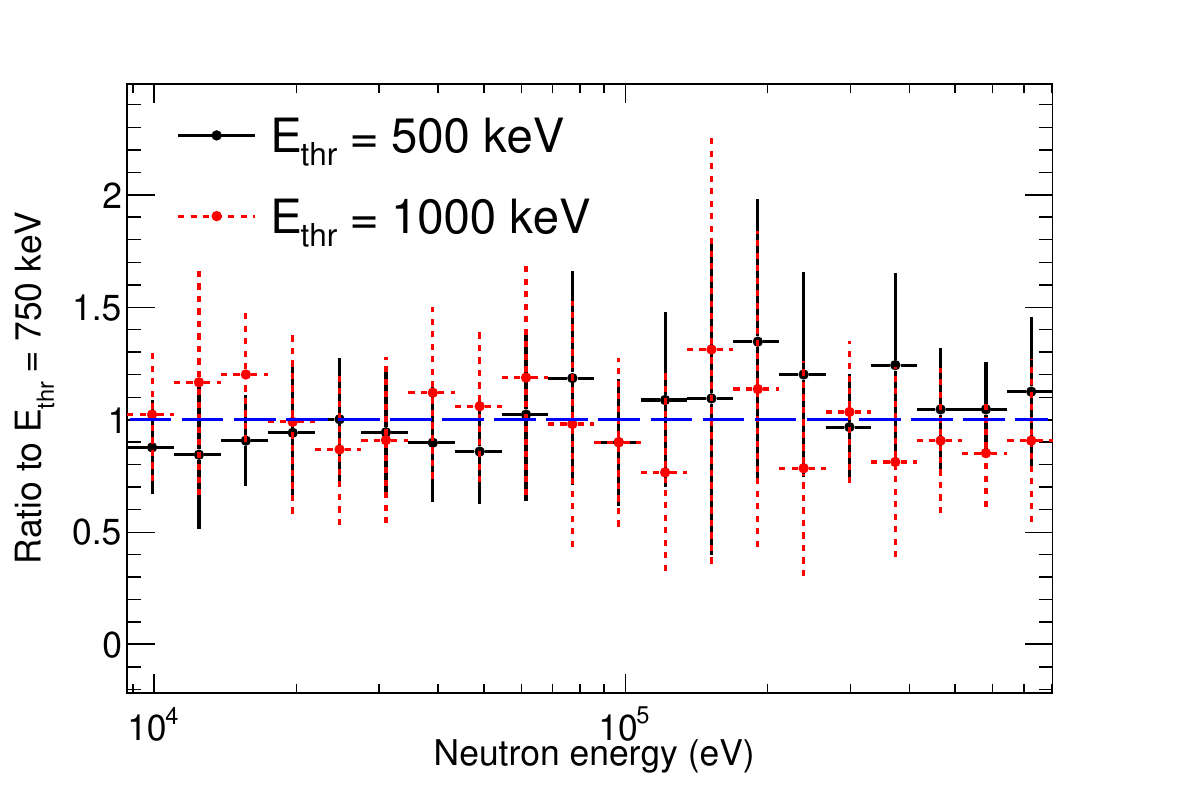}
\caption{Ratio of the extracted $^{242}Pu$ capture count rate with three different thresholds in $\gamma$-ray deposited energy $E_{thr}=500, 750$, and 1000~keV for E$_n\geq$10~keV.}
\label{Fig_FissionSubtracted}
\end{center}
\end{figure}

The fission contribution was the reason behind the large discrepancies between the results obtained for different $\gamma$-ray thresholds $E_{thr}$ above $E_n$=100~keV, shown in Fig.~\ref{Fig_ValidScalingGammas}. 
After correcting the counting rates for this contribution, the results for $E_{thr}\gtrsim$500~keV show a good agreement within the statistical uncertainties (see Fig.~\ref{Fig_FissionSubtracted}). 
%However, a systematic deviation of about 8\% is found in the ratios of Fig.~\ref{Fig_FissionSubtracted}, which may be related to the uncertainty in the fission correction. Indeed, the observed deviation is at the level of the systematic uncertainty associated with the fission cross section in that neutron energy range.

According to the discussion carried out throughout this section, the neutron energy range above 100~keV can be analyzed when a sufficiently high $E_{thr}\gtrsim 750$~keV is applied.
The negligible impact of the $\gamma$-flash in the fast C$_{6}$D$_{6}$ detectors combined with the rejection of the ($n$,$n'$) and correction for the ($n$,$f$) backgrounds, allows us to report capture data up to a neutron energy of about 600~keV. Beyond this energy, the 
steep increase in systematic uncertainty (15--20\%) associated with the ($n$,$f$) (see  bottom Fig.~\ref{Fig_FissionCaptureSimulations}) leads to a total systematic uncertainty clearly above the target accuracy.

\subsection{\label{sec:Yield} Capture yield: normalization and corrections}
The neutron capture yield  $Y$ is defined as the probability for an incident neutron to undergo a capture reaction and is related to the capture $\sigma_{\gamma}$
and total $\sigma_{tot}$ cross sections as:
\begin{equation}
Y= (1-e^{-n\sigma_{tot}})\frac{\sigma_{\gamma}}{\sigma_{tot}},
\label{eq:Yieldtheor} 
\end{equation}
where all the quantities but $n$ are neutron energy dependent.

Experimentally, the $^{242}$Pu capture yield in the URR was determined using the following expression:
 \begin{equation} 
Y(E_{n})= F_{norm}^{thr}\frac{C(E_{n})-B(E_{n})}{\Phi(E_{n})\cdot\varepsilon_{c}(E_{n})}, 
\label{eq:Yieldexp} 
 \end{equation}
$C(E_{n})$ and $B(E_{n})$ being the unweighted distributions of total and background counts per pulse displayed in Fig.~\ref{Fig_summary_bckgs}, 
$\Phi(E_{n})$ the total number of neutrons of a given energy $E_{n}$ reaching n\_TOF-EAR1 in each pulse~\cite{Barbagallo:2013} and
$\varepsilon_{c}(E_{n})$ the detection efficiency, which has been assessed with the methodology discussed in Sec.~\ref{sec:Efficiency}. Last, $F_{norm}^{thr}$
is the threshold-dependent normalization factor calculated as:
 \begin{equation} 
 F_{norm}^{thr}= \frac{\langle W^{thr}\rangle\cdot F^{thr}_{c}}{f^{thr}_{SRM}}, 
\label{eq:YieldNorm} 
 \end{equation}
where $F_{c}^{thr}$ is the product of $f_{mc}$, $f_{thr}$ and $f_{ce}$, correction factors to the efficiency due to multiple counting, 
fraction of the response below the detection threshold and internal conversion.
The reader is referred to Ref.~\cite{JLerendegui:RRR} for a detailed
description of these corrections and associated uncertainties. $f^{thr}_{SRM}$ is the absolute normalization factor obtained via the SRM~\cite{SRM_Macklin} by fitting the 
plateau of the saturated resonance at 4.9~eV in the capture yield of a thick (100~$\mu$m) $^{197}$Au target with the SAMMY code~\cite{SAMMY}. 
The average weighting factor $\langle W^{thr}\rangle$ was introduced in Sec. II.B (see also Ref.~\cite{JLerendegui:RRR}). The use of the AWF method allows one to assess the absolute capture efficiency at low neutron energies. The absolute normalization $F_{norm}^{thr}$ has been calculated from the RRR ($E_n\leq$1~keV) only for $E_{thr}$ =150~keV used to analyze the neutron-energy range $E_n<$100~keV. The yield at $E_n>$100~keV, analyzed with a threshold $E_{thr}$ =750~keV, has been normalized using the 2.67 eV resonance. The total systematic uncertainty related to the normalization is slightly below 2\%, as described in detail in Ref.~\cite{JLerendegui:RRR}. 

%The normalization factor $F_{c}^{thr}$ was actually
%calculated just for the lowest threshold used in this analysis, $E_{thr}$ = 150~keV to minimize the correction factor $F_{c}^{150~keV}$ = 1.035(3). 
%The first resonance in the normalized yield obtained with $E_{thr}$ = 150~keV was used to normalize the unweighted yields obtained with higher detection thresholds. 

%The saturated resonance of $^{197}$Au has been fitted for each detector to extract individual normalization factors f$_{SRM}$, as listed in Table~\ref{Table_SRM}. 
%Figure \ref{Fig_AuNormSAMMY} shows the fit of this resonance in the average yield of the four detectors using the SAMMY code \cite{SAMMY}. 

\section{\label{sec:Results} Results}
\begin{figure}[h]
\begin{center}
\includegraphics[width=9.0cm]{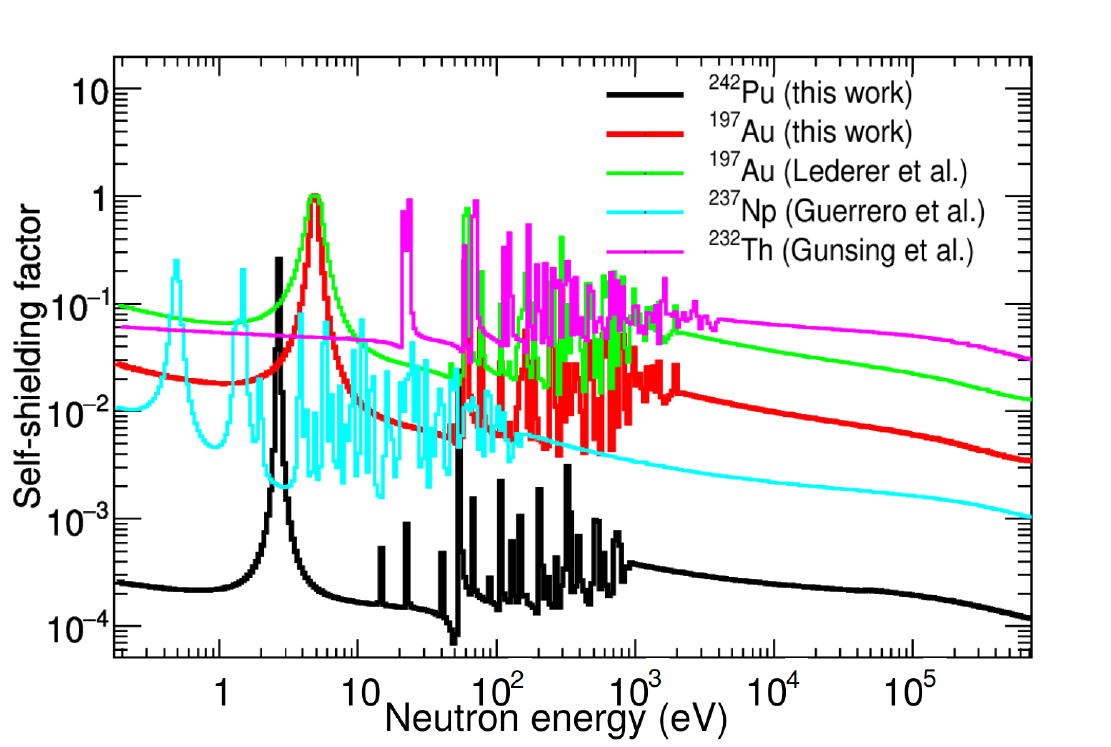}
\caption{Neutron self-shielding correction factor (1-e$^{-n\sigma_{tot}}$) as a function of neutron energy for the two samples measured in this work $^{197}$Au and $^{242}$Pu compared to previous published data at n\_TOF coming from Lederer~\cite{Lederer:2011}, Guerrero~\cite{237Np} and Gunsing~\cite{Gunsing:2006}.}
\label{Fig_Selfshielding}
\end{center}
\end{figure}

%\subsection{\label{sec:crosssection} Average capture cross section in the unresolved resonance region}
%\begin{itemize}
%\item from capture yield to cross section: Thin-target approximation. Highlight again the advantage of a thin target in terms of negligibe multiple scattering corrections.
%\end{itemize}
In the unresolved resonance region the averaged capture cross section in a given neutron energy interval, or bin, is calculated using the thin target approximation as:
\begin{equation}
\langle\sigma_{\gamma}(E_{n})\rangle_{URR} = \frac{\langle Y (E_{n}) \rangle}{n}.
\label{eq:AvgXSection} 
\end{equation}
This approximation is fully justified as the self-shielding and multiple scattering corrections can be neglected due to the characteristics of the $^{242}$Pu and $^{197}$Au targets. The latter was used to check the applied procedure (see Sec.~\ref{sec:Aucrosssection}). This is illustrated in Fig.~\ref{Fig_Selfshielding}, where the self-shielding correction factor given by (1-e$^{-n\sigma_{tot}}$) for the targets used in this work is compared to those of previous n\_TOF measurements of Au and other actinides using thicker targets~\cite{Gunsing:2006,Lederer:2011,237Np}. The corrections in the energy region of interest (above 1~keV) is of the 
order of 1\% for $^{197}$Au and negligible for $^{242}$Pu, which inherently implies also a negligible multiple scattering correction.

%The cross section in the resonance region was parameterized in terms of individual resonances from the experimental capture yield by means of the R-Matrix formalism~\cite{JLerendegui:RRR}.
%On the contrary, in the unresolved resonance region, the 
%capture yield calculated using Eqs.~(\ref{eq:Yieldexp}-\ref{eq:YieldNorm}) is actually an average yield due to the overlapping resonances. The average yield $\langle Y (E_{n}) \rangle$, 
%is related to the average capture cross section $\langle \sigma_{\gamma} \rangle$ by

%where $F_{sample}(E_{n})$ is a sample-related correction for self-absorption and multiple scattering. In Fig.~\ref{Fig_Selfshielding}, the self-absorption correction for the two samples in this work is compared to previous capture measurements carried out
%at n\_TOF-EAR1 covering the~keV region. The largest self-shielding factors in the URR in Fig.~\ref{Fig_Selfshielding} correspond tomeasurements on $^{232}$Th~\cite{Gunsing:2006} and $^{197}$Au~\cite{Lederer:2011}, for which $F_{sample}(E_{n})$ 
%was calculated with help of SESH or MCNP.
%On the other hand, in the analysis of the measurement on $^{237}$Np~\cite{237Np} it was proved that a small self-absorption factor (0.5\% or smaller according to Fig.~\ref{Fig_Selfshielding}) is associated with a negligible multiple scattering correction.  
%Thus, we can conclude that the use of thin $^{242}$Pu targets in this measurement, leading to a self-absorption correction below 0.05\% in the URR,allows us to apply the thin-target approximation (i.e. $F_{sample}(E_{n})$~=~1).

\subsection{\label{sec:Aucrosssection} Validation of the measurement and analysis in the URR: $^{197}$Au($n$,$\gamma$)}
The $^{197}$Au($n$,$\gamma$) ancillary measurement aims mainly at obtaining the absolute normalization of the capture yield (see Eq.~(\ref{eq:Yieldexp})); additionally, we took advantage of the fact that the $^{197}$Au($n$,$\gamma$) cross section 
is very accurately known in the URR~\cite{IAEA:Standards,Lederer:2011,Massimi:2014} to validate
the measurement and analysis techniques used in the $^{242}$Pu, since the full 
data reduction process (without a correction for fission) followed the same steps for both $^{197}$Au and $^{242}$Pu.

\begin{figure}[!h]
\begin{center}
\includegraphics[width=8.0cm]{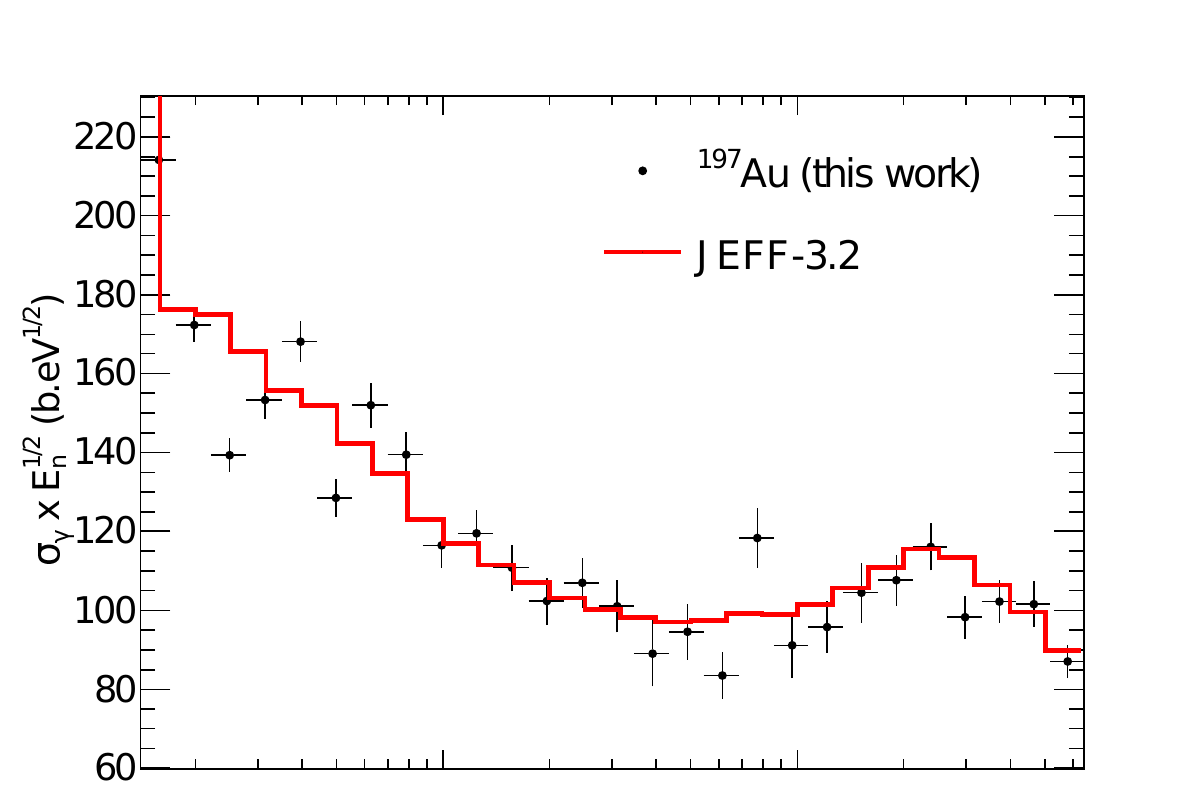}
\includegraphics[width=8.0cm]{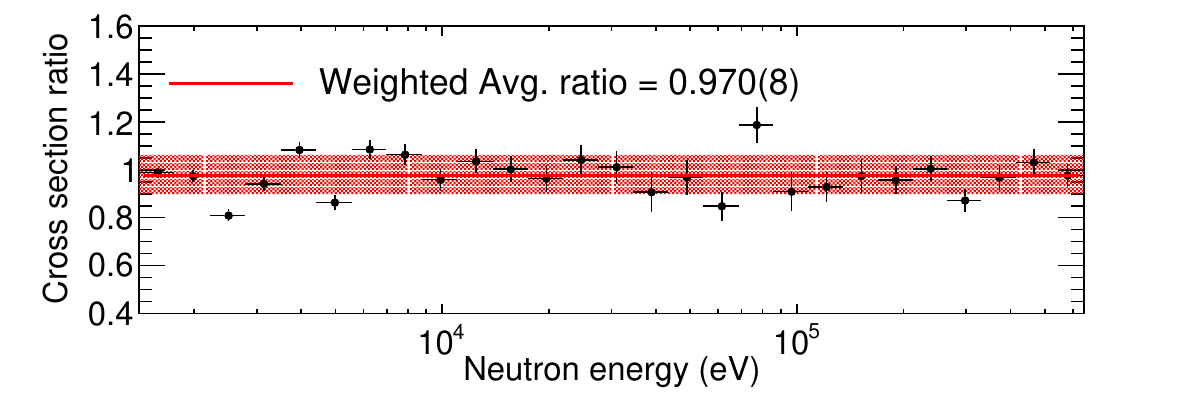}
\caption{Top: Comparison of the capture cross section of $^{197}$Au in the URR obtained in this work with JEFF-3.3. Bottom: Ratio of this work with respect to JEFF-3.3 showing an average ratio of 0.970(8) and a good agreement up to a neutron energy of 600~keV. The shadowed corridor corresponds to the standard deviation of the ratio.}
\label{Fig_Au_Xsection}
\end{center}
\end{figure}

The resulting capture cross section of $^{197}$Au in the URR with a $\gamma$-ray energy threshold $E_{thr}$ of 250~keV is compared to JEFF-3.3 (=JEFF-3.3) in Fig.~\ref{Fig_Au_Xsection};
the cross sections in other evaluated libraries and the IAEA Standard~\cite{IAEA:Standards} are fully consistent with JEFF. The top panel of Fig.~\ref{Fig_Au_Xsection} 
shows the capture cross section multiplied by $E_{n}^{1/2}$ for plotting purposes. The measured and evaluated cross sections agree on average within 3\% and the ratio fluctuates around this value with a standard deviation of 7\% (shadowed corridor in the bottom panel of Fig.~\ref{Fig_Au_Xsection}), consistent with the statistical uncertainties. The slightly larger discrepancies observed below 5~keV are related to the resonant structures in the 
$^{197}$Au cross section. At higher energies the largest differences are obtained at around 80~keV, which corresponds to the Al absorption dip of the n\_TOF neutron 
flux. The comparison stops at 600~keV, above which the inelastic channels of $^{197}$Au start playing a role (see Ref.~\cite{Lederer:2011} for details).

%The largest discrepancies are found in the energy region below 5~keV and could be a consequence of the presence of resonant structures, not included in the smooth evaluated cross section.
%Above this energy just one energy bin, corresponding
%to the flux dip at 80~keV deviates more than 2$\sigma$ from the evaluated cross section. The last data point in the comparison corresponds to a neutron energy just above 600~keV at which the contribution of the inelastic
%channels starts to be present, as it was shown in a dedicated measurement of this cross section at n\_TOF-EAR1~\cite{Lederer:2011}.

\subsection{\label{sec:Pucrosssection} $^{242}$Pu capture cross section in the URR }

\begin{table*}[!htb]
\begin{center}
\caption{Summary of individual relative systematic uncertainties in each source (second column) and their propagated contribution to the total systematic uncertainty in the capture cross section. The URR has been divided into four neutron energy ranges (columns 3-6) to evaluate the energy dependence of these uncertainties. The maximum, minimum and average systematic uncertainty $\Delta_{syst}$ (squared sum of individual contributions) and the integrated relative statistical uncertainty $\Delta_{stat}$ in each energy range are presented at the end of the table.}
\begin{tabular}{lccccc} 
\hline \hline
                             &                             & \multicolumn{4}{c}{\textbf{Uncertainty propagated to the cross section(\%)}}  \\
\hline
\textbf{Sources uncertainty} &  \textbf{Unc. (\%)} & \textbf{1--10~keV} & \textbf{10--50~keV}  & \textbf{50--250~keV}   & \textbf{250--600~keV}  \\
\hline
\textbf{$E_{n}$-independent}    \\
%\hline

~Normalization yield                   &         2           &          2          &         2          &    2       &        2              \\
~Target mass                           &         4           &          4          &         4          &    4       &        4               \\
%\hline
\textbf{$E_{n}$-dependent}\\
%\hline
~Efficiency                  &         3--3.5           &      3              &        3           &    3--3.5       &        3.5             \\
~Dummy                               &        1            &        5.2         &       6.3          &    6.9          &         4.0      \\
~Beam-off                            &        1            &        0.8         &       0.5          &    0.3          &         0.1           \\
~In-beam $\gamma$-rays               &        $^a$         &         3          &       3            &    3           &          3          \\  
~Scattered neutrons                  &        $^a$         &        1.5         &        1           &    2.5          &        1.5       \\
~Fission     		             &  0.5--6.5           &         0        &       0.5          &     1.4         &        6.8         \\
~Neutron flux                        &        2--5          &        2           &       4--5          &   2--5 $^b$   &            2          \\ 
\hline
\textbf{Range of $\Delta_{syst}$}    &                  & \textbf{8.4--8.7}     &\textbf{9.7--10.1}  &\textbf{9.4--10.2} &     \textbf{9.0--12.3} \\
\textbf{Average $\Delta_{syst}$}    &                      & \textbf{8.5}        &\textbf{9.9}      &\textbf{10.0} &     \textbf{10.4}       \\
\textbf{Integrated $\Delta_{stat}$}    &                     &\textbf{2.9}        &\textbf{5.8}       &\textbf{13.0}  &  \textbf{11.9}        \\

\hline \hline    
\end{tabular}
\label{Table_Uncertainties}
\end{center}
\textit{a) The resulting systematic uncertainty was estimated from the extracted yield and not propagated from the uncertainty of the individual background components (see text for details).
\\b) Uncertainty in the neutron flux: 4--5\% (50--100~keV),$\sim$2\% (100--250~keV).}
\end{table*}

\begin{figure}[!h]
\begin{center}
\includegraphics[width=8.5cm]{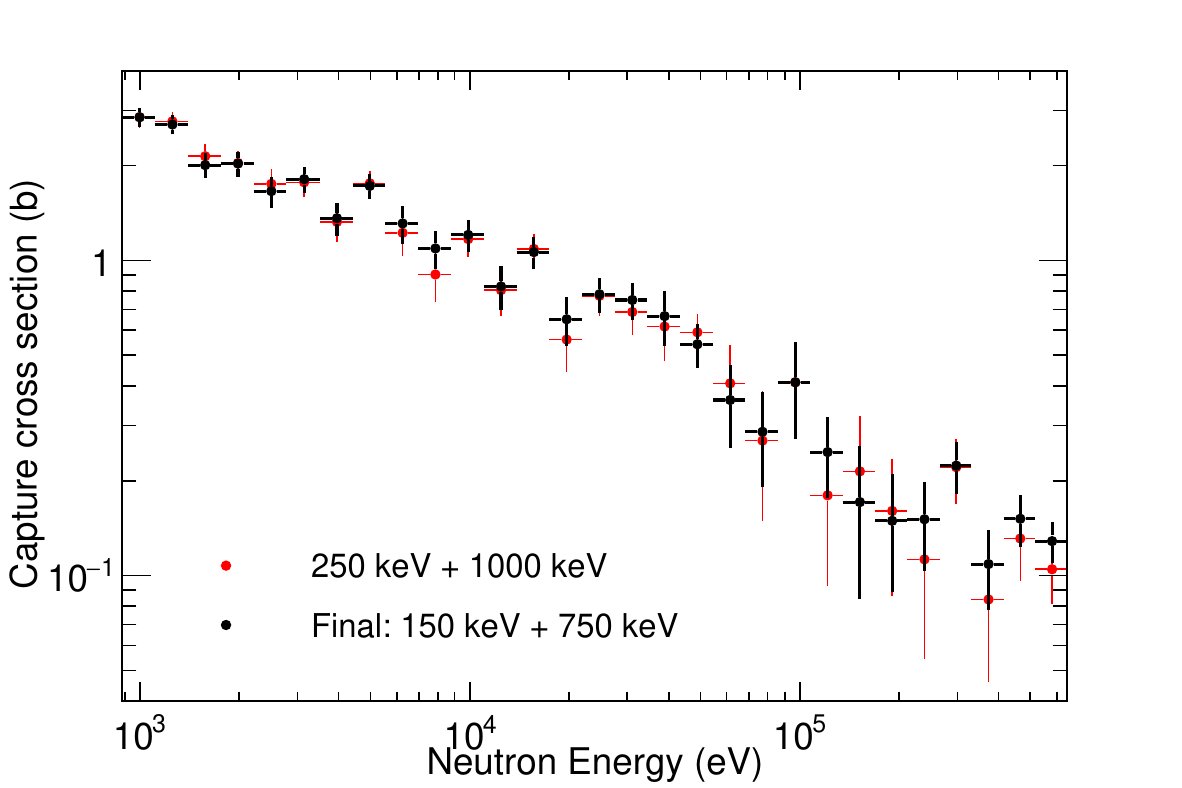}
\caption{Final $^{242}$Pu capture cross section obtained in this work (black) obtained with $E_{thr}$ = 150~keV below $E_n$=100~keV and 750~keV at higher neutron energies. For comparison we show also results using higher thresholds ($E_{thr}$ = 250~keV and 1000~keV, in red). The results are fully consistent with both threshold sets. The displayed uncertainties are only statistical.}
\label{Fig_Pu242_XSection_Thresholds}
\end{center}
\end{figure}

Following the data reduction process described in Sec.~\ref{sec:Analysis}, the final $^{242}$Pu capture cross section in the URR was obtained 
using two detection thresholds for low and high neutron energy. As shown in Fig.~\ref{Fig_Pu242_XSection_Thresholds} the results for $E_n<$100~keV do not depend on the tested threshold (150 or 250~keV) and thus $E_{thr}=150~keV$, the lowest has been chosen in order to keep the statistics as high as possible. For neutron energies above 100~keV the fission
and inelastic background contributions become more important and hence a minimum $\gamma$-ray energy threshold of 750~keV has been chosen to minimize the fission contribution and suppress the inelastic background. This $E_{thr}$ has been used for the analysis of this neutron energy region, as shown in Fig.~\ref{Fig_Pu242_XSection_Thresholds}. The final averaged capture cross section with approximately 10 bins per decade is then also listed in Table~\ref{Table_Xsection} together with the
statistical and systematic uncertainties.

%different detection thresholds. The final result
%was built by combining different thresholds in the following way.
%For neutron energies from 1 to 100~keV, the different values of $E_{thr}$ provide consistent results as it is shown in Fig.~\ref{Fig_ValidScalingGammas}.
%However, the statistical uncertainty is significantly enhanced with increase threshold and, thus, a low detection threshold ($E_{thr}\leq$~250~keV) is used for the final
%cross section below $E_{n}$~=~100~keV. At higher energies, a threshold
%higher than 500~keV is required to obtain compatible results. 

The individual sources of systematic uncertainties are summarized in Table~\ref{Table_Uncertainties}, leading to the total systematic uncertainties in the cross section listed in Table~\ref{Table_Xsection}. First, we give those that do not depend on the neutron
energy, related to the normalization ($2\%$) and $^{242}$Pu mass ($4\%$). These sources of uncertainty are common to the results of the resonance region~\cite{JLerendegui:RRR}
since they represent an uncertainty in a global normalization. 
On the other hand, the neutron-energy-dependent uncertainties, related to the background subtraction 
and the neutron flux, are the major contributions to the total systematic uncertainty in the cross section in the URR and have been evaluated
in four energy intervals. The systematic uncertainty in the 
dummy background, estimated to be 1\%, dominates the overall uncertainty in the cross section below 250~keV as it accounts for more than 75\% of the counts.
The uncertainty in this background is obtained by combining a 0.5\% associated with the beam monitoring (see Sec.~\ref{sec:Detection}) with
an additional 0.5\% related to an estimated uncertainty of up to 10\% in the backing layer thickness, as it was discussed in Sec.~\ref{sec:Backgrounds}. We would like to comment on the uncertainty associated with the efficiency (PHWT) that was determined by combining the uncertainty in the average weighting factor (2\%), extracted in analysis of the resonance region~\cite{JLerendegui:RRR},  with the (1--1.5\%) systematic uncertainty coming from the simulation-based assessment of the neutron-energy dependency of the efficiency, as detailed in Sec.~\ref{sec:Efficiency}.

\begin{table}[!b]
\begin{center}
\caption{Averaged neutron capture cross section $\langle\sigma_{\gamma}\rangle$, absolute uncertainties $u_{stat}$ and $u_{syst}$ and relative $\Delta_{stat}$ and $\Delta_{syst}$ (in \%) for each neutron energy bin. See Table~\ref{Table_Uncertainties} for the contribution of the individual sources to the systematic uncertainty.}
\begin{tabular}{cccccc} 
\hline \hline
\textbf{$E_{low}$} &  \textbf{$E_{high}$} & \textbf{$\langle\sigma_{\gamma}\rangle\pm u_{stat} \pm u_{syst}$}  & \textbf{$\Delta_{stat}$} & \textbf{$\Delta_{syst}$}  \\
    \textbf{(keV)}       &  \textbf{(keV)}      &  \textbf{(b)}                                                &      \textbf{(\%)}       &      \textbf{(\%)}             \\
\hline	
0.883	&	1.111	&	2.85	$\pm$	0.19	$\pm$	0.24	&	6.8	&	8.4	\\
1.111	&	1.398	&	2.71	$\pm$	0.19	$\pm$	0.23	&	6.9	&	8.4	\\
1.398	&	1.759	&	2.01	$\pm$	0.19	$\pm$	0.17	&	9.2	&	8.4	\\
1.759	&	2.214	&	2.03	$\pm$	0.18	$\pm$	0.17	&	9.1	&	8.4	\\
2.214	&	2.785	&	1.66	$\pm$	0.19	$\pm$	0.14	&	12	&	8.5	\\
2.785	&	3.504	&	1.81	$\pm$	0.17	$\pm$	0.15	&	9.6	&	8.5	\\
3.504	&	4.408	&	1.36	$\pm$	0.17	$\pm$	0.12	&	12	&	8.5	\\
4.408	&	5.544	&	1.73	$\pm$	0.16	$\pm$	0.15	&	9.1	&	8.5	\\
5.544	&	6.973	&	1.31	$\pm$	0.18	$\pm$	0.11	&	14	&	8.6	\\
6.973	&	8.769	&	1.09	$\pm$	0.15	$\pm$	0.09	&	14	&	8.6	\\
8.769	&	11.03	&	1.21	$\pm$	0.14	$\pm$	0.11	&	11	&	8.7	\\
11.03	&	13.86	&	0.83	$\pm$	0.13	$\pm$	0.08	&	16	&	9.7	\\
13.86	&	17.42	&	1.07	$\pm$	0.12	$\pm$	0.10	&	12	&	9.8	\\
17.42	&	21.90	&	0.65	$\pm$	0.11	$\pm$	0.06	&	18	&	9.9	\\
21.90	&	27.51	&	0.79	$\pm$	0.10	$\pm$	0.08	&	13	&	10	\\
27.51	&	34.56	&	0.76	$\pm$	0.10	$\pm$	0.08	&	14	&	10	\\
34.56	&	43.41	&	0.68	$\pm$	0.13	$\pm$	0.07	&	20	&	10	\\
43.41	&	54.50	&	0.55	$\pm$	0.09	$\pm$	0.05	&	16	&	10	\\
54.50	&	68.40	&	0.36	$\pm$	0.11	$\pm$	0.04	&	30	&	10	\\
68.40   &	85.82	&	0.29	$\pm$	0.10	$\pm$	0.03	&	34	&	10	\\
85.82	&	107.6	&	0.42	$\pm$	0.14	$\pm$	0.04	&	34	&	10	\\
107.6	&	134.9	&	0.245	$\pm$	0.071	$\pm$	0.024	&	29	&	9.9	\\
134.9	&	169.0   &	0.170	$\pm$	0.062	$\pm$	0.016	&	36	&	9.7	\\
169.0   &	211.7	&	0.148	$\pm$	0.048	$\pm$	0.014	&	32	&	9.4	\\
211.7	&	264.9	&	0.150	$\pm$	0.047	$\pm$	0.014	&	31	&	9.4	\\
264.9	&	331.3	&	0.219	$\pm$	0.042	$\pm$	0.020	&	19	&	9.0	\\
331.3	&	414.0	&	0.109	$\pm$	0.031	$\pm$	0.010	&	29	&	9.2	\\
414.0   &	516.9	&	0.151	$\pm$	0.029	$\pm$	0.017	&	19	&	11	\\
516.9	&	645.1	&	0.128	$\pm$	0.019	$\pm$	0.015	&	15	&	12	\\

\hline \hline    
\end{tabular} 
\label{Table_Xsection}
\end{center}
\end{table}

An additional relevant contribution to the uncertainty is the 
energy dependence of the neutron flux, that is known with an accuracy ranging from 2\% up to 4--5\% in the energy region of the dips (10--100~keV)~\cite{Barbagallo:2013}. Last, the uncertainty associated with the correction for the fission background, shown in Fig.~\ref{Fig_FissionCaptureSimulations}, has been evaluated individually for each energy bin in Table~\ref{Table_Xsection}. 
Only the average uncertainty due to the fission background over wider energy ranges is then given in Table~\ref{Table_Uncertainties}. Evidently, this is the major source of uncertainty above $E_n=250$~keV. The contribution of the inelastic background is removed by using a high $\gamma$-ray energy threshold and no associated uncertainty is considered.

The average total systematic uncertainty in each of wide energy intervals in Table~\ref{Table_Uncertainties} ranges from
8\% below 10~keV to 10\% above 50~keV. Using the binning presented in Table~\ref{Table_Xsection}, the statistical uncertainties are larger than the systematic ones already from about 1.5~keV. However, if the cross section is integrated over the energy ranges in Table~\ref{Table_Uncertainties}, the statistical uncertainty remains below or similar the systematic one. 
We conclude that considering only the systematic uncertainty we have achieved the desired accuracy of 8--12\% in the averaged cross section required 
for all the fast reactors listed in Table~\ref{Table_summary} from 1 to 600~keV. If we take into account the statistical uncertainty, very wide bins need to be used to reach the aimed accuracy for neutron energies above about 50~keV, compare Tables~\ref{Table_Uncertainties}~and~\ref{Table_Xsection}. 

%From the average values of these uncertainties in the energy intervals in Table~\ref{Table_Uncertainties}, we have interpolated the value for each neutron energy bin.

\subsection{\label{sec:FITACS} Average parameter description of the cross section }

\begin{figure}[!t]
\begin{center}
\includegraphics[width=8.5cm]{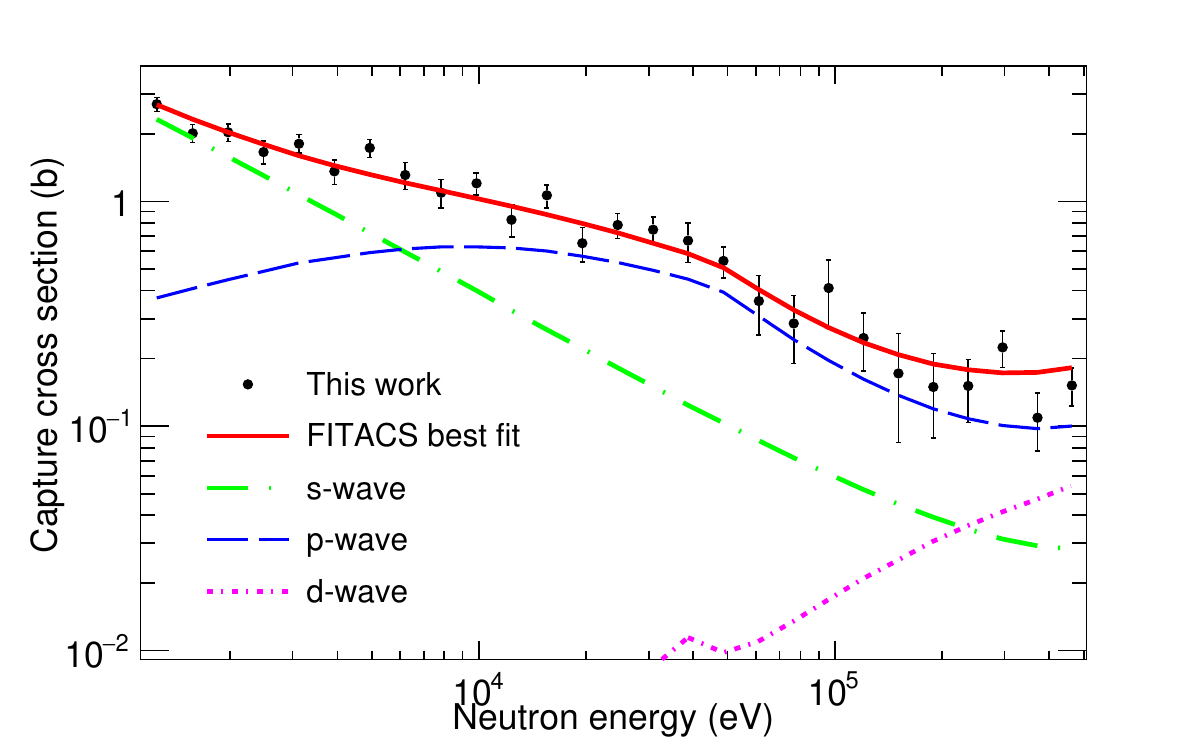}
\caption{Capture cross section in the URR obtained in this work together with the fit from FITACS. The error bars correspond only to the statistical uncertainties. The contribution of the different angular momentum components coming from the fit is included; the fitted parameters are listed in Table~\ref{Table_ParametersFitacsRRR}.}
%Bottom: Relative contribution (\%) of the s-, p- and d-waves.}}
\label{Fig_Pu242_XSection_FITACS}
\end{center}
\end{figure}

The final experimental $^{242}$Pu capture cross section between 1 and 600~keV, shown in Figs.~\ref{Fig_Pu242_XSection_Thresholds} and ~\ref{Fig_Pu242_XSection_FITACS},
has been fitted using the Hauser-Feshbach approach~\cite{Froehner:1989} to get average resonance parameters.
%This description is specially relevant since it is the usual approach to generate resonances in the calculation of self-shielding factors or probability tables. 
To do that we have used the FITACS code~\cite{Froehner:1989} included in SAMMY. The average resonance parameters to be varied in this code are the neutron strength functions $S_{\ell}$ and average radiation widths~$\langle\Gamma_{\gamma}\rangle_{\ell}$ different neutron orbital momenta $\ell$. We used a fixed resonance spacing $D_0$ = 15.8 eV (that then also determines $D_\ell$ for higher $\ell$~\cite{SAMMY}). Further, FITACS assumes that the average radiative widths are the same for all resonances with the same parity, i.e. $\langle\Gamma_\gamma\rangle_{\ell=0} = \langle\Gamma_\gamma\rangle_{\ell=2}$ in our case. As a result, only $S_\ell$ for three $l$ values and two independent $\langle\Gamma_\gamma\rangle_{\ell}$ were fitted. The channel radius was assumed to be $R^\prime=9.75$ fm~\cite{Mughabab:06}, and the impact of the neutron inelastic scattering was considered using energy, spin and parity of the low-lying $^{242}$Pu levels from ENSDF~\cite{ENSDF}.

%The average radiative width is considered to be only parity-dependent ($\langle\Gamma_{\gamma}\rangle_{\ell=0}=\langle\Gamma_{\gamma}\rangle_{\ell=2}\neq \langle\Gamma_{\gamma}\rangle_{\ell=1}$).The average level spacing is given just for s-wave resonances $D_{0}$. The average spacing for higher angular momenta is calculated assuming a $2J+1$ dependence for the level density.The input for SAMMY includes also the distant level parameters $R^{\infty}_{\ell}$ and the nuclear radius $a~=~A^{1/3}+0.8 fm = 8.46~fm$, which are related to the scattering radius $R'~=~9.75~fm$ by
%\begin{equation}
%R'=~a\cdot(1-R^{\infty}).
%\end{equation}
%The energy dependence of the parameters as well as the impact of the neutron inelastic scattering are also considered in the SAMMY/FITACS calculation. For this purpose, the energy, spin and parity of the $^{242}$Pu levels were extracted from ENSDF~\cite{ENSDF}.

The cross section fit from FITACS is plotted in Fig.~\ref{Fig_Pu242_XSection_FITACS} together with the individual contribution from $\ell=0,1,2$. The average resonance parameters obtained from the fit are then listed in Table~\ref{Table_ParametersFitacsRRR} together with the $s$-wave average parameters obtained in the analysis of the resolved resonance region~\cite{JLerendegui:RRR}. The fitted value of $s$-wave parameters perfectly agree with that deduced from the RRR, where resonances were analyzed up to $E_n$=4~keV~\cite{JLerendegui:RRR}. This agreement confirms the consistency of the cross sections extracted in the RRR and URR below a few~keV, where resonances with $\ell=0$ dominate. At higher neutron energy, the contribution from $\ell>0$ clearly dominates the capture cross section, see Fig.~\ref{Fig_Pu242_XSection_FITACS}, and the fitting is virtually insensitive to the s-wave parameters.

\begin{table}[!htb]
\begin{center}
\caption{Average resonance parameters -- neutron strength functions $S_\ell$, average radiative widths $\langle \Gamma_\gamma\rangle_\ell$, and resonance spacings $D_\ell$ -- obtained from the fit of the capture cross section from this work between 1 and 600~keV using the FITACS code, see Fig.~\ref{Fig_Pu242_XSection_FITACS} for the fit. The average resonance parameters obtained from the statistical analysis of the resolved resonance region~\cite{JLerendegui:RRR} are also presented for comparison. Values of $D_\ell$ were fixed during fitting, see text for details.}
\begin{tabular}{cccc} 
\hline \hline
\textbf{Angular momentum} &  \textbf{$S_{\ell}$ x 10$^{4}$}    & \textbf{$\langle\Gamma_{\gamma}\rangle_{\ell}$ (meV)}  & \textbf{$D_{\ell}$ (eV)}\\
\hline
  $s$-wave ($\ell=0$)       &    0.90(12)    &  24.1(20)    &   15.8       \\
  $p$-wave ($\ell=1$)       &    2.6(3)      &   28(3)    &   5.35       \\
  $d$-wave ($\ell=2$)       &    0.6(3)     &   24.1(20)    &   3.30       \\
 \hline
 $s$-wave (RRR~\cite{JLerendegui:RRR})   &   0.91(8)   &  24.8(5)  &  15.8(8)     \\
\hline \hline    
\end{tabular} 
\label{Table_ParametersFitacsRRR}
\end{center}
\end{table}

\subsection{\label{sec:Comparison} Comparison to existing data and evaluations }

\begin{figure}[!htb]
\begin{center}
\includegraphics[width=8.4cm]{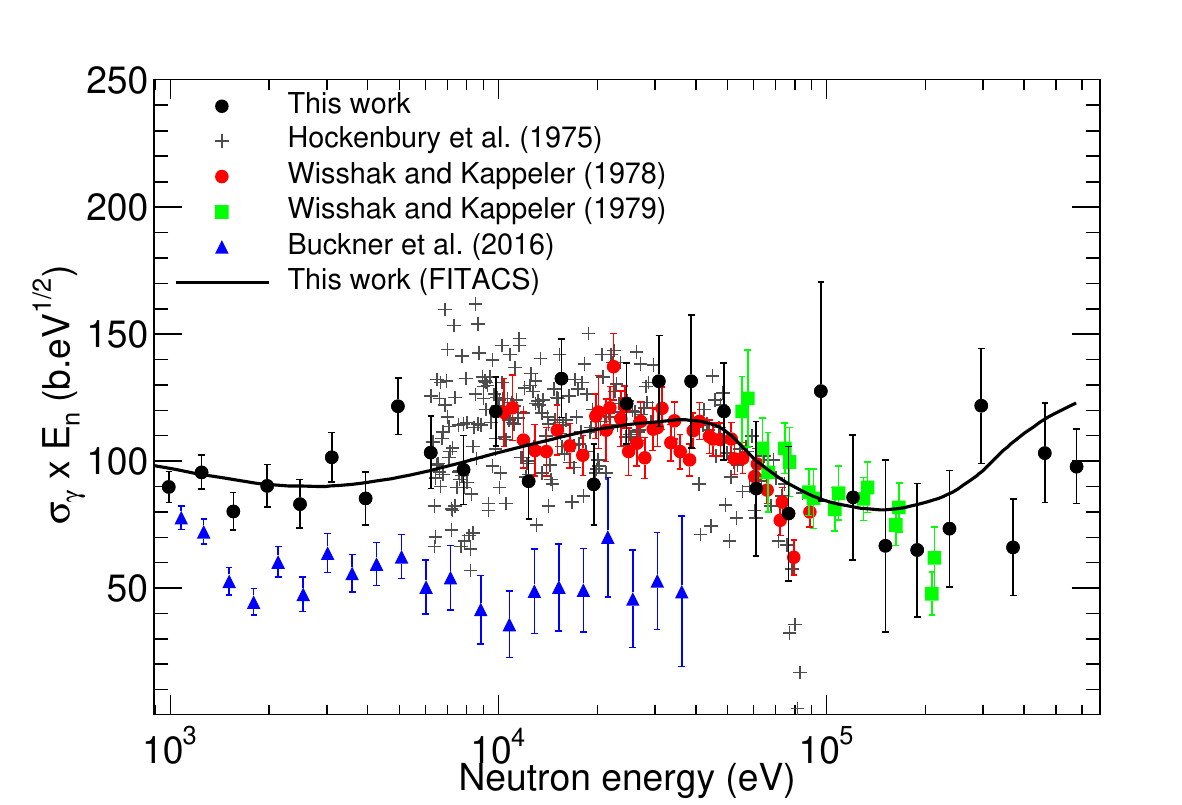}
\includegraphics[width=8.4cm]{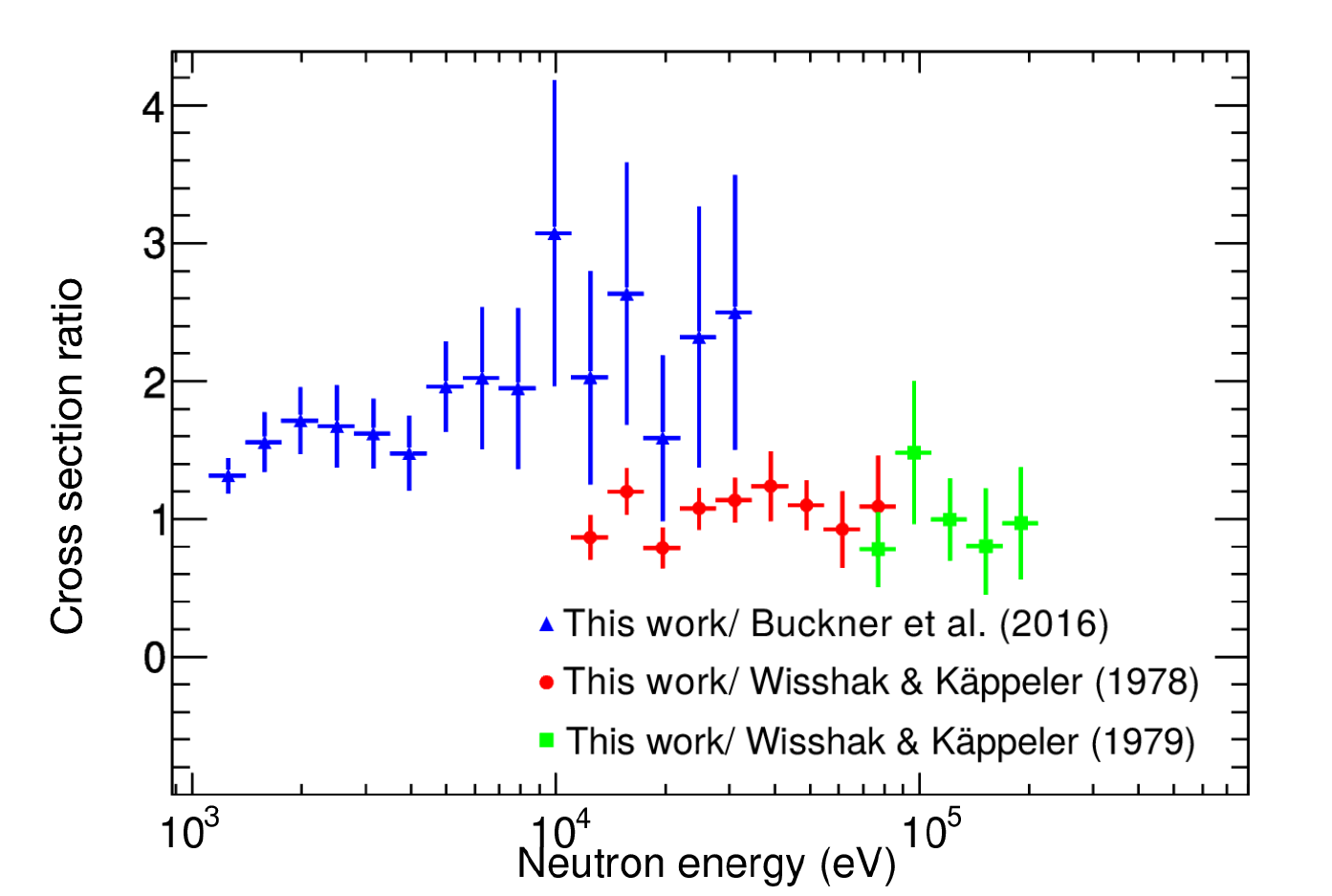}

\caption{Top: Capture cross section of $^{242}$Pu in the URR obtained in this work together with the fit from FITACS compared to the previous measurements available in EXFOR. The cross section has been multiplied by $\sqrt{E_n}$. Only statistical uncertainties are displayed. Bottom: Ratio of this work with respect to the previous ones. The ratio to the data by Hockenbury et al. was not included due to its large dispersion and the absence of error bars.}
\label{Fig_Pu242_vsPrevious}
\end{center}
\end{figure}

The capture cross section reported in this work becomes the first data set to cover in a single measurement the complete energy range from 1 to 600~keV.
The upper panel of Fig.~\ref{Fig_Pu242_vsPrevious} compares our results with the recent measurement from 1 to 40~keV at LANSCE by Buckner et al.~\cite{Buckner:2016} and with the three time-of-flight measurements from the 70's by
Hockenbury at RPI (6--87~keV)~\cite{Hockenbury:1975} and Wisshak and K\"appeler at FZK (10--90 and 50--250~keV)~\cite{Wisshak_Kaeppeler:1978,Wisshak_Kaeppeler:1979}. 
The cross sections by Wisshak and K\"appeler in Fig.~\ref{Fig_Pu242_vsPrevious} have been calculated via multiplication of the experimental $^{242}$Pu($n$,$\gamma$)/$^{197}$Au($n$,$\gamma$) 
data given in Refs.~\cite{Wisshak_Kaeppeler:1978,Wisshak_Kaeppeler:1979} by the $^{197}$Au($n$,$\gamma$) cross section in JEFF-3.3. Our cross section is in agreement with the two data sets of Wisshak and K\"appeler within uncertainties. On the other hand, it is systematically above the recent measurement at LANSCE (Buckner et al.) at all presented energies. As evident from Fig.~\ref{Fig_Pu242_vsPrevious} the difference is in all cases at least about one, but typically 2-3 standard deviations and the cross section ratio can be up to a factor of 3. This deviation is remarkable when compared to the results in the RRR, where our resonance kernels below 500~eV were found to be in average only 6\% larger~\cite{JLerendegui:RRR}. The inconsistency between the RRR and URR data of LANSCE may arise from the sharp deterioration of the signal-to-background ratio beyond 1~keV~\cite{Buckner:2016}. Last, the ratio to the measurement by Hockenbury et al. is not presented since the data points are widely spread and their uncertainties are not reported.

\begin{figure}[!t]
\begin{center}
\includegraphics[width=8.5cm]{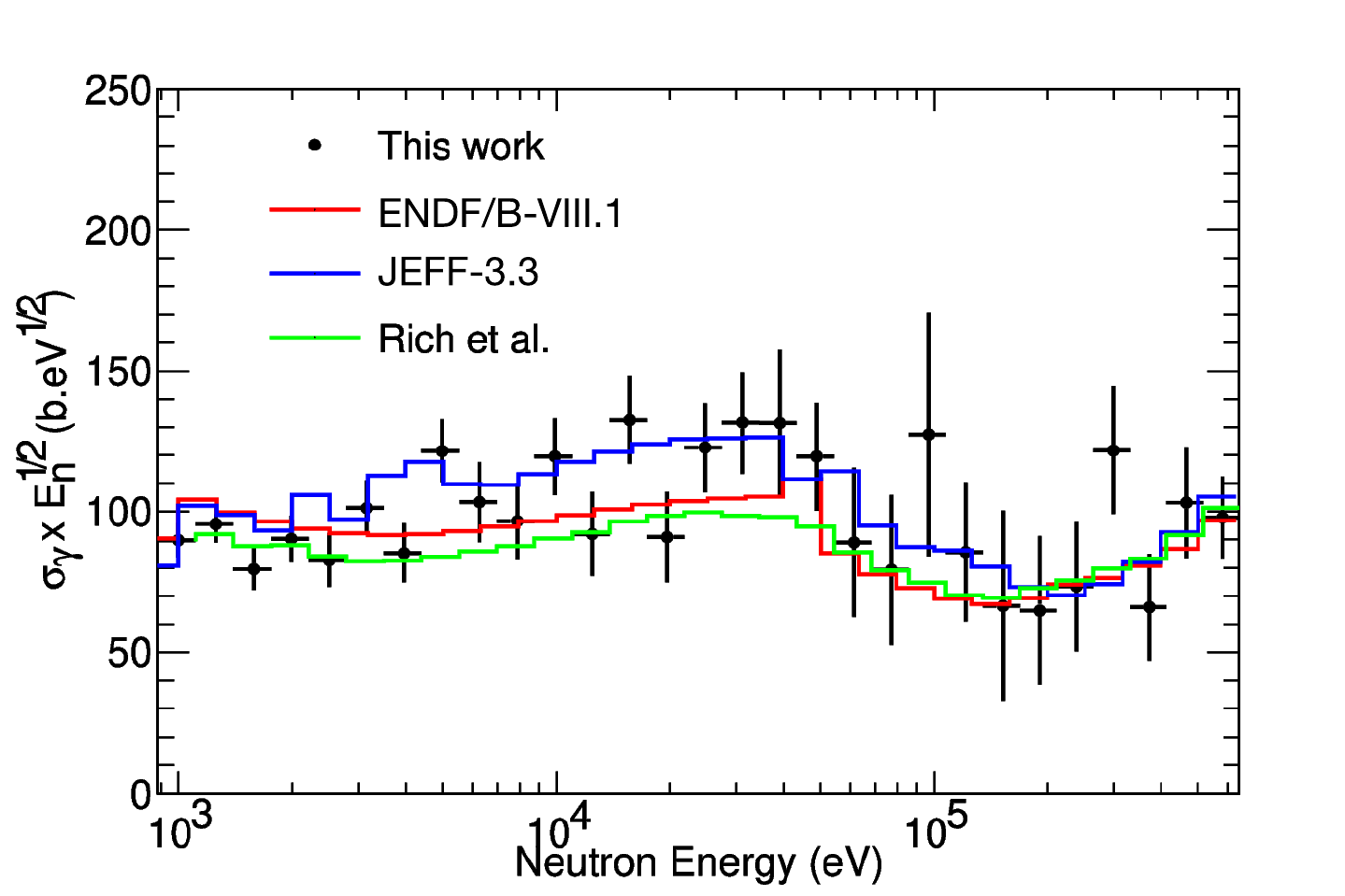}
\includegraphics[width=8.65cm]{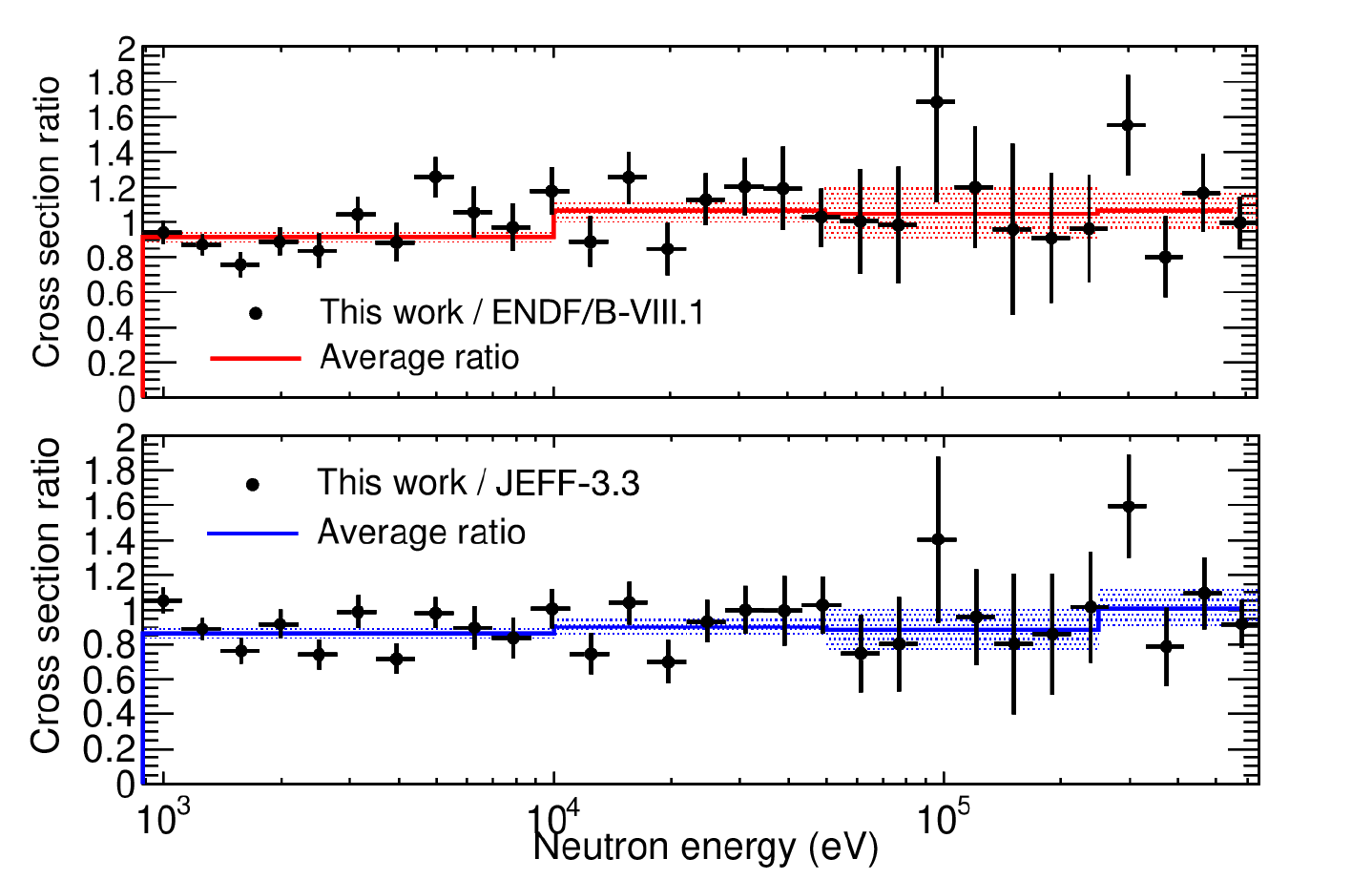}
%\includesvg[width=8.5cm]{Fig16A_Pu242_URR_Xsection_multSqrtEn_nTOF_vs_Evaluations_RICH_linScale_afterFissionSubt_axis_updatedLibraries.svg}
%\includesvg[width=8.65cm]{Fig16B_Pu242_URR_Ratio_evaluations_AvgRatio_shadowedUnc_4Ranges_upto600keV_afterFissionSubt_updatedLibraries.svg}
\caption{Top: Comparison of the capture cross section of this work with the main evaluated files, JEFF-3.3 (blue) and ENDF/B-VIII.1 (red). Only statistical uncertainties are displayed. The cross section calculated by Rich et al.~\cite{RichNSE162} is also shown in green. Bottom: Ratio between the cross section of this work and the evaluations. The solid lines correspond to the average ratios in four energy ranges and the shadowed corridors to the statistical uncertainty of the average 
(results in Table~\ref{Table_RatioEvaluation}).}
\label{Fig_Pu242_vsEvaluations}
\end{center}
\end{figure} 

The top panel of Fig.~\ref{Fig_Pu242_vsEvaluations} then shows the capture cross section of this work compared to the ENDF/B-VIII.1~\cite{ENDF} and JEFF-3.3~\cite{JEFF} evaluated files. 
The cross section calculated by Rich et al., aiming at describing
all the neutron induced reactions on $^{242}$Pu with a set of consistent average resonance parameters~\cite{RichNSE162}, is also included as a reference.

\begin{table*}[htb!]
\begin{center}
\caption{Average ratio of the cross section of this work with respect to JEFF-3.3 and ENDF/B-VIII.1 
in different energy ranges. The ratio is also indicated in the bottom panel of Fig.~\ref{Fig_Pu242_vsEvaluations}. The uncertainties in brackets are only statistical. The systematic ones are given
in the last row.}
\begin{tabular}{lcccc} 
\hline \hline
                             &  \textbf{1--10~keV} & \textbf{10--50~keV}  & \textbf{50--250~keV} &  \textbf{250--600~keV}\\
\hline
ENDF/B-VIII.1                &      0.92(3)      &        1.06(6)       &       1.04(13)       &  1.06(11)   \\
JEFF-3.3                    &      0.86(3)      &        0.90(6)       &       0.88(11)       &  1.00(10) \\
\hline
Systematic uncertainty (\%)  &       8.5          &	  9.9	        &       10.0           &    10.4       \\

%Rich et al.                 && \textbf{8.1}        &\textbf{9.5}       &\textbf{$<$11$^b$}\\ &                  &                         &                      \\
\hline \hline    
\end{tabular} 
\label{Table_RatioEvaluation}
\end{center}
\end{table*}

The ratio of this work with respect to the evaluations is shown in the bottom panel of the same figure and in Table~\ref{Table_RatioEvaluation}. Both in the figure and the table, the average ratios for four wide neutron energy ranges between 1 and 600~keV are shown. The ratios illustrate the differences between the libraries and indicate that our cross section is significantly lower than both JEFF-3.3 and ENDF/B-VIII.1 below $E_n$=10~keV. When comparing with JEFF-3.3, our cross section is about 10--14\% below the evaluation between 1 and 250~keV. This is important because it could at least partially explain the overestimation of about 14\% in the calculated over experimental (C/E) ratios for the capture rate on $^{242}$Pu samples when interpreting the PROFIL and PROFIL-2 data with JEFF-3.3. At higher energies, an overall agreement is found between our data and evaluated cross sections within the systematic uncertainty.

\section{\label{sec:Summary} Summary and conclusions}
The neutron capture cross section of $^{242}$Pu has been measured by means of the time-of-flight technique at the CERN n\_TOF-EAR1 facility employing an array of four 
C$_{6}$D$_{6}$ scintillators as a detection system.
The target consisted of seven thin layers of $^{242}$Pu enriched to 99.959\%, each 45~mm in diameter, with a total mass of 95(4)~mg electrodeposited on thin aluminum backings. 
The main advantages of using a series of thin fission-like $^{242}$Pu targets for a capture measurement has been highlighted throughout the paper.

The full capture data measured at n\_TOF-EAR1 covers the neutron energy interval from 0.1 eV to a few hundreds of~keV. 
A previous publication on this measurement presented the analysis of the resolved resonance region up to 4~keV~\cite{JLerendegui:RRR},
focusing on the \textit{R}-Matrix resonance analysis and statistical analysis of the $s$-wave resonance parameters. This paper deals with the analysis of the high energy data, focusing on the extraction of the cross section in the unresolved resonance region (URR). A detailed description of the analysis technique, background subtraction, efficiency estimation and limiting factors to expand the upper energy limit has been presented in this paper. We determined the cross section up to 600~keV. This energy limit comes from the increasing systematic uncertainty associated with the correction of the fission background at higher energies. The new capture data thus covers the required energy range for all the nuclear innovative systems proposed in Ref.~\cite{WPEC26}.

The final systematic uncertainties of the measured cross section below 250~keV range from about 8\% to 10\%, achieving the accuracy required by the NEA WPEC-26 group listed in Table~\ref{Table_summary}, and are mainly dominated
by the uncertainty in the beam-related background that contributes more than 75\% of the total count rate in the $^{242}$Pu measurement in this neutron energy range.
Above 250~keV the systematic uncertainty ranges from 10 to 12\% and is dominated by the fission background contribution. While the systematic uncertainty affects the global accuracy of the cross section, 
the statistical uncertainty dominates above a few~keV if relatively narrow energy intervals are used (as those in Table~\ref{Table_Xsection}).

The capture cross section obtained in this work is the first one to cover the energy range from 1 to 250~keV in a single measurement and the first capture measurement beyond 250~keV. Our results show a good agreement with
the two measurements by Wisshak and K\"appeler for neutron energies between 10 and 250~keV and are at variance with the latest LANSCE measurement. The comparison with the evaluations indicates a 10--14\% smaller average capture cross section compared
to JEFF-3.3 in the energy range from 1 to 250~keV and a good agreement above this energy, which helps to achieve consistency between the integral experiments and the cross section data.

Last, the measured cross section was used to determine average resonance parameters by means of a Hauser-Feshbach calculation with the SAMMY/FITACS code. The fitted values of $S_{0}$ and $\langle\Gamma_{\gamma}\rangle_{0}$ are consistent with those extracted from the statistical analysis of the resonances below 4~keV. 
%On the other hand, the parametrized cross section suggests an increase in the partial p-wave cross section (given by $S_{1}$ and $\langle\Gamma_{\gamma}\rangle_{1}$) compared to the values in the literature. 
%It is worth mentioning that the accuracy of the fitted parameters is limited by the sizable correlation among them and the lack of total cross section data to constrain the fit.
%We can conclude that we are in fairly good agreement with JEFF-3.3 up to 1~keV.

In summary, the new $^{242}$Pu capture data at n\_TOF-EAR1 provide the first data set in the URR covering in a single measurement the full energy range up to  600~keV and supports the trend indicated by the 
PROFIL experiments to reduce the capture cross section in JEFF-3.3. This work complements the previously published data in the resonance region~\cite{JLerendegui:RRR} and at thermal~\cite{JLerendegui:Thermal} and shall lead to a consistent re-evaluation of both the resolved and unresolved resonance regions.

\section*{Acknowledgments}
This measurement has received funding from the EC FP7 Programme under the projects NEUTANDALUS 
(Grant No. 334315) and CHANDA (Grant No. 605203), the Spanish Ministry of Economy and Competitiveness projects FPA2013-45083-P, FPA2014-53290-C2-2-P and FPA2016-77689-C2-1-R
and the V Plan Propio de Investigaci\'on Programme from the University of Sevilla.  Additionally, we thank the support provided by the postdoctoral grant CIAPOS/2022/020 funded by the Generalitat Valenciana and the European Social Fund. Support from the German Federal Ministry for Education and Research (BMBF), contract number 03NUK13A,
and the Croatian Science Foundation under the project IP-2022-10-3878, as well as from the funding agencies of all other participating institutes are are gratefully acknowledged.

\end{document}